\documentclass[onecolumn]{aastex6}

\collaborationName{Friends of AASTeX}
\usepackage{booktabs}

\def\msun{{\rm\,M_\odot}}
\def\lsun{{\rm\,L_\odot}}

\def\s-1{{\rm\,s^{-1}}}

\def\spose#1{\hbox to 0pt{#1\hss}}
\def\C3H2{{\rm\,\rm C_3H_2}}
\def\NH3{{\rm\,\rm NH_3}}
\def\HOCO+{{\rm\,\rm HOCO^+}}
\def\lta{\mathrel{\spose{\lower 3pt\hbox{$\mathchar"218$}}
     \raise 2.0pt\hbox{$\mathchar"13C$}}}
\def\gta{\mathrel{\spose{\lower 3pt\hbox{$\mathchar"218$}}
     \raise 2.0pt\hbox{$\mathchar"13E$}}}

\def\lsim{\raise0.3ex\hbox{$<$}\kern-0.75em{\lower0.65ex\hbox{$\sim$}}} 
\def\gsim{\raise0.3ex\hbox{$>$}\kern-0.75em{\lower0.65ex\hbox{$\sim$}}} 

\begin{document}


\title{High-resolution ALMA Study  of the Proto-Brown-Dwarf Candidate L328-IRS}



\author{Chang Won Lee\altaffilmark{1,2}, Gwanjeong Kim\altaffilmark{1,2,3},  Philip C. Myers\altaffilmark{4},  Masao Saito\altaffilmark{5}, Shinyoung  Kim\altaffilmark{1,2}, Woojin Kwon\altaffilmark{1,2}, A-Ran Lyo\altaffilmark{1}, Archana Soam\altaffilmark{1}, \& Mi-Ryang Kim\altaffilmark{1} }
\altaffiltext{1}{Korea Astronomy \& Space Science Institute, 776 Daedeokdae-ro, Yuseong-gu, Daejeon, Republic of Korea.  E-mail: cwl@kasi.re.kr}
\altaffiltext{2}{Korea University of Science \& Technology, 217 Gajungro, Yuseong-gu, Daejeon, 305-333, Republic of Korea.}
\altaffiltext{3}{Nobeyama Radio Observatory, National Astronomical Observatory of Japan, National Institutes of Natural Sciences, Nobeyama, Minami- maki, Minamisaku, Nagano 384-1305, Japan}
\altaffiltext{4}{Harvard-Smithsonian Center for Astrophysics, 60 Garden Street, Cambridge, MA  02138, USA.}
\altaffiltext{5}{National Astronomical Observatory of Japan, National Institutes of Natural Sciences, 2-21-1 Osawa, Mitaka, Tokyo 181-8588, Japan}

\begin{abstract}

This paper presents our observational attempts to precisely measure the central mass of a proto-brown dwarf candidate, L328-IRS, in order to investigate whether L328-IRS is in the substellar mass regime.
Observations were made for the central region of L328-IRS with the dust continuum  and CO isotopologue line emission at ALMA band 6,
discovering  the detailed outflow activities and a deconvolved disk structure of  a size of $\sim$87 AU  $\times$ 37 AU.  We investigated the rotational velocities as a function of the disk radius, finding that its motions between 130 AU and 60 AU are partially fitted 
with  a Keplerian orbit  by a stellar object of $\sim0.30 ~\msun$, while the motions within 60 AU do not follow  any Keplerian orbit at all. This makes it difficult to lead a reliable estimation of the mass of L328-IRS.
Nonetheless, our ALMA observations were useful enough to well constrain the  inclination angle of the outflow cavity of L328-IRS as  $\sim 66^{\circ}$, enabling us to better determine 
the mass accretion rate of $\sim 8.9\times 10^{-7} \msun~yr^{-1}$. From assumptions that the internal luminosity of L328-IRS is mostly due to this mass accretion process in the disk, or that L328-IRS has mostly accumulated the mass through this constant accretion rate during its outflow activity,  its mass was estimated to be $\sim 0.012 - 0.023\msun$, suggesting L328-IRS to be a substellar object.
However, we leave our identification of L328-IRS as a proto-brown dwarf to be tentative because of various uncertainties especially regarding the mass accretion rate. 

\end{abstract}




\section{Introduction} \label{sec:intro}
The formation of brown dwarfs (BDs) is not yet well understood and still under strong debate.
There are many suggested ideas on how objects acquire the BD mass, including 
turbulent or gravitational fragmentation in a molecular cloud, like the case of low mass star formation (Padoan \& Nordlund 2004; Bonnell et al. 2008; Bate 2012), 
ejections of substellar objects  from multiple star-forming regions or  massive circumstellar disks  (Basu \& Vorobyov 2012; Boss 2001; Reipurth \& Clarke 2001; Bate 2009),
tidal shear and high velocity dispersion present in a stellar cluster (Bonnell et al. 2008), and photo-erosion of a prestellar core by a nearby OB star (Whitworth \& Zinnecker 2004). 
The underlying principle in these mechanisms is mostly related to the premature termination of accreting processes 
toward the point sources (see Luhman 2012; Lee et al. 2013).

A problem here  is that  such formation scenarios are mostly theoretical and hardly tested.
Various observations on the early precursors of BDs like the proto-BDs containing initial conditions of the BD formation will be essential for unraveling the mystery of BD formation. 

In practice, however, identifying a bona-fide proto-BD is very difficult. For this work we need to know the envelope mass,  the mass accretion rate, and the central mass of the candidate.
The envelope mass is  important  as it determines a mass to supply to the central source. The mass accretion rate is needed to estimate the mass to be accreted to the central object during 
the main accretion phase.  
The mass of the central object is the most essential as it makes possible to determine the final mass of the central source when it is added to the mass to be accreted during its remained main accretion phase. 

The envelope mass is rather simply derived using the continuum observations. However,  estimating the accretion rate and the central mass is not trivial, requiring detailed observations of the outflow and disk structures  and
 suffering from many uncertain parameters and/or observational difficulties.
For example, there are two best candidates of the proto-BD, IC348-SMM2E (Palau et al. 2014) and L328-IRS (Lee et al. 2013).  However, their identification as the proto-BD is yet to be concluded.
IC348-SMM2E has been suggested to be a strong candidate of a proto-BD from its low  envelope mass ($ \sim 0.03 \msun$), small accretion rate ($\sim 1.6\times 10^{-7} \msun yr^{-1}$), and its low central mass
($ \sim 0.02 \msun$) from the SMA observations (Palau et al. 2014). However, their estimation of accretion rate was obtained by assuming an isothermal accretion process (Shu 1977) and 
is not yet confirmed by observations. Moreover the central mass was roughly inferred by assuming that the disk-like structure follows the Keplerian motions without any further detailed kinematical analysis of their SMA data. 

L328-IRS has a similar level of confidence about its BD status. It was also suggested to be a potential candidate of the proto-BD because it has small values for the envelope mass ($\sim 0.09~\msun$), the accretion rate ($\sim 3.6\times 10^{-7} \msun yr^{-1}$), and 
 the central mass ($\sim 0.05~\msun$) (Lee et al. 2013).
In this case, its envelope mass was fairly well determined from the continuum observations.  However, the mass accretion rate was estimated from CO outflow observations with many uncertain parameters, 
especially the unknown inclination angle of the outflow. Moreover, the kinematical estimation of the central mass of this source was not observationally possible because of 
the absence of  spectral line observations  in its possible disk region in high angular resolution. Instead  we estimated its total accreted mass and assumed it to be the mass of the central source.  
 
In recent ALMA era, however, it now becomes  possible  for us to look at very detailed outflow and disk structures at once in a protostellar or proto-BD system. 
This helps to constrain many parameters regarding these structures, and especially make a kinematical diagnosis of the disk rotation 
over the central object in a scale of a few tens AU  to enable to directly estimate the mass of the central object in its disk. 

In this paper we present our ALMA observations of L328-IRS region with sub-arc second angular resolution as well as  large recoverable scale of a few tens arc second, 
aiming to identify an accreting disk in  L328-IRS and constrain the detailed outflow structures, respectively.
From this study we attempt to make a direct estimation of  the mass of the central object as well as  a more accurate calculation of the mass accretion rate in order to confirm whether L328-IRS belongs to the regime of a substellar mass.  

L328-IRS was first identified  in a dense core L328 as a Very Low Luminosity Object (VeLLO) from  observations by the $Spitzer$ Space Telescope (Lee et al. 2009). 
Its bipolar outflows in a pc scale were found from single dish observations (Lee et al. 2013). The mass accretion rate inferred from the outflow property was found to be smaller 
by  almost one order of magnitude than the canonical accretion rate for low-mass protostars. Considering that its envelope mass is small, it has been suggested to be a strong precursor of the brown dwarf 
unless it has  accreted enough mass in the past to evolve to a normal solar-type star. 
Because of inward motions found in global scale over its envelope
and some hint that L328-IRS is forming  through a gravitational fragmentation process of the envelope, L328-IRS has been regarded as one of the best examples supporting an idea that a brown dwarf forms like a normal low-mass star
(Lee et al. 2013).

In the next section we describe ALMA observations toward L328-IRS in continuum and CO isotopolgue lines.
Then we explain results seen in the continuum and molecular line observations in section 3.
In section 4 we discuss possible Keplerianity in the disk rotation, and how L328-IRS system would look in 3D space to help to reliably constrain an inclination angle of the outflow cavity from L328-IRS,
the mass accretion rate, and the central mass of L328-IRS.  
We summarize our results  in the last section.

\section{Observations} \label{sec:Obs}

L328-IRS was observed with ALMA  in $^{12}$CO, $^{13}$CO, C$^{18}$O 2-1 lines, and 1.3 mm continuum during two observing dates in  two different configurations of Cycle 2 campaign (with an ALMA project number of 2013.1.00783.S).

The first observation was made on September  1st, 2014 in C34-6 configuration with its maximum  baseline of about 1091m 
to achieve an angular resolution of $0\arcsec.33  \times 0\arcsec.25$ at 230 GHz.
The second observation was carried out on April 14, 2015 in C34-2 configuration. Its maximum  baseline is  about 305m giving an angular resolution of $1\arcsec.43 \times 0\arcsec.78$
while its minimum baseline  is about 14.1m putting a limit of a maximum recoverable scale of $11\arcsec.4$.  
In both observations 34 (12-m) antennas were used.

We combined the data from two configuration observations in visibility space and deconvolved the combined data to obtain the final image data.
All the observing parameters are summarized in Table 1.
We reduced our data with different  robust weights 1) Natural, 2) Uniform, 3) long baseline data only (baselines $\rm > 300 k\lambda$), and 4) r (robust weight) =0.5 to find any possible detailed structure in an angular resolution as high as possible.
We found that the central disk-like emission is not bright enough to be useful for diagnosing its detailed kinematics with long baseline data only, for example, the robust weighted long baseline data (baselines $\rm > 300 k\lambda$).  
Among the data with different robust weights that we tested, the data with  r=0.5 gave the most reasonable quality for our forthcoming analysis in terms of 
sensitivity and spatial resolution. The synthesized beam size of our combined data with r=0.5 is 0{\arcsec}.31$\times$0{\arcsec}.23 (P.A.=79{$^{\circ}.3$}) in continuum emission.  
The 1 $\sigma$ sensitivities  for continuum and line emission are measured to be $\rm \sim0.04~mJy~beam^{-1}$ and $\rm \sim4.0~mJy~beam^{-1}~[0.096~km/s]^{-1}$, respectively (Table 1).

\setcounter{table}{0}
\begin{table}[h!]
\centering
\rotate
\caption{Summary of Observing parameters} \label{tab:tbl1}
\begin{tabular}{ccccc}
\tablewidth{0pt}
\hline
\hline
Target	&	\multicolumn{4}{c}{L328-IRS}	\\
\cmidrule(r){1-5}
Continuum Center Coordinate	&	\multicolumn{4}{c}{($\alpha, \delta$)$_{J2000}$=($\rm
18^h16^m59\fs4973\pm0.0002^s, -18^{\circ} 02\arcmin 30\arcsec.263\pm 0.001\arcsec$)}	\\
\cmidrule(r){2-5} 
Observing dates	&	\multicolumn{2}{c}{Sept. 1, 2014}	& \multicolumn{2}{c}{April 14, 2015} \\
\cmidrule(r){2-3}
\cmidrule(r){4-5}
ALMA Configuration	&	\multicolumn{2}{c}{C34-6}	&	\multicolumn{2}{c}{C34-2} \\
Number of antenna	&	\multicolumn{2}{c}{34}	&	\multicolumn{2}{c}{34} \\
Maximum baseline		&	\multicolumn{2}{c}{1091.0 m} 	&	\multicolumn{2}{c}{304.6 m}   \\
Minimum baseline		&	\multicolumn{2}{c}{40.6 m} 	&	\multicolumn{2}{c}{14.1 m}  \\
Bandpass calibrator	&	\multicolumn{2}{c}{J1733-1304} 	&	\multicolumn{2}{c}{J1733-1304}  \\
Flux calibrator	&	\multicolumn{2}{c}{Neptune} 	&\multicolumn{2}{c}{Neptune}	\\
Gain calibrator	&	\multicolumn{2}{c}{J1733-1304} 	&\multicolumn{2}{c}{J1733-1304}	\\

\cmidrule(r){2-5}
Observations		&	Continuum	&	$^{12}$CO2-1	&	$^{13}$CO2-1		&	C$^{18}$O2-1	\\
\cmidrule(r){2-5}
Frequencies	& 231.5922 GHz	&	230.5380 GHz	&	220.3987	GHz&	219.56036 GHz \\
\cmidrule(r){2-2}\cmidrule(r){3-5}
Synthesized beam of combined data	&0{\arcsec}.31$\times$0{\arcsec}.23(P.A.=79{$^{\circ}.3$})	&		&	0{\arcsec}.33$\times$0{\arcsec}.25(P.A.=76{$^{\circ}$.9})	&	 \\
RMS of combined data	&	$\sim$0.04 mJy~beam$^{-1}$	&	\multicolumn{3}{c}{$\sim$4.0 mJy~beam$^{-1}$~(0.096 km/s)$^{-1}$} \\ 
\hline
\end{tabular}
\end{table}

\section{Results} \label{sec:results}

\subsection{1.3mm continuum}

A single compact source of a ``disk-like'' structure at L328-IRS position was detected in 1.3 mm continuum emission  within a primary beam field of view of $\sim 25{\arcsec}.2$ (Figure 1).

This continuum source was marginally resolved with  the synthesized beam size. 
The image component size in FWHM of the continuum is about $0\arcsec .54~(\pm 0\arcsec.01)\times  0\arcsec.33(\pm 0\arcsec.004)$ (P.A. $\sim 101^{\circ}.2\pm 0^{\circ}$.8). 
Its  FWHM size  deconvolved by the beam is about  $0\arcsec.40 ~(\pm  0\arcsec.01) \times 0\arcsec.17 ~(\pm 0\arcsec.01) $  (P.A. $\sim 106^{\circ}.5\pm 1^{\circ}.6$) which corresponds to $\sim$87 AU  $\times$ 37 AU
in a linear size at its distance ($\sim 217$ pc, Maheswar et al. 2011).
Peak position of the continuum is measured from its 2D Gaussian fit as ($\alpha, \delta$)$_{J2000}$=($\rm
18^h16^m59\fs4973\pm0^s.0002, -18^{\circ} 02\arcmin 30\arcsec.263\pm 0\arcsec.001$).
The total integrated and peak intensities are estimated to be $\rm 11.40~(\pm 0.22)$ mJy  and  $\rm 6.43~(\pm 0.09)$ mJy $\rm beam^{-1}$, respectively.
We note that there is no other emission structure except for this disk-like compact emission in the field of view. 

The mass for the continuum was calculated from 
$$
\rm M_{cont} = {{S_{1.3mm} D^2} \over {B(T_{dust}) \kappa_{1.3mm}}},
\eqno (1)
$$
 where $S_{1.3mm}$ is the flux density at 1.3 $mm$, D is the distance of L328-IRS  from us, 
$\rm B(T_{dust})$ is the Planck function at a dust temperature ($\rm T_{dust}$), and
$\kappa_{1.3mm}$ is the dust opacity at 1.3 mm.  If $\kappa_{1.3mm}$ is assumed to be $\kappa_\lambda =0.1\times ({{0.3~mm} \over {\lambda}})^\beta$(Beckwith et al. 1990), 
then  $\kappa_{1.3mm}$ is given as $\rm \kappa_{1.3mm}=0.014~cm^{2}~g^{-1}$ with $\beta=1.3$ (Chandler \& Sargent 1993). A gas to dust mass ratio is assumed to 
be 100 in converting the dust continuum mass into the total mass of the continuum source. 

Our estimated total mass for the disk structure is about 0.01 $\rm M_\odot$  at the dust temperature of 16 K  from Lee et al. (2013) with a measured flux density of 11.4 mJy. 
 However, note that the dust temperature of 16 K  is the temperature for the dust envelope of 
about $\sim 20\arcsec$ ($\sim 4300$ AU) size surrounding L328-IRS and thus  the temperature of the compact continuum source very near to L328-IRS can be higher. Thus its likely mass may be  smaller than 0.01$\rm M_\odot$
as estimated here.
For example, if the temperature is assumed to be as high ($\sim 30$ K) as 
that of B335 in smaller scale ($\sim 600$ AU) (e.g., Yen et al. 2015), the continuum mass would be as small as about 0.0045 $\rm M_\odot$.
 
The most interesting feature in the continuum emission is its disk-like shape whose position coincides with the location of  L328-IRS. We note that its  position angle (P.A. $\sim 106^{\circ}.5\pm 1^{\circ}.6$) is significantly different from that of  the beam shape by about $27^{\circ}$,
meaning that  the disk-like shape emission is not likely affected by the observed beam, but is instead a real feature. 

 \subsection{$\rm CO$ isotopologue lines emission}
 \subsubsection{Distribution of CO emission}
CO 2-1 line emission is known to well trace various kinematical structures of the molecular outflows in embedded YSOs (e.g., Bontempt et al. 1996).  Previous single dish CO 2-1 observations by Lee et al. (2013) have shown that  L328-IRS is producing 
CO outflows  in a bipolar shape of the blue and red components in a sub-parsec scale.
Our ALMA CO observations are found to well delineate the outflow part very near to L328-IRS  in the range of $\rm \sim 0 ~km~s^{-1}$ and $\rm 14.0~km~s^{-1}$ in a consistent manner with the result by the single dish observations, but with many more detailed structures of the compact outflow close to L328-IRS.
Figure 2 shows the velocity channel maps of CO emission, displaying various outflow-related  structures near to L328-IRS. 
One interesting structure that can be noticed is a rather high velocity component seen only toward the central region of the disk-like continuum emission, probably emanating from L328-IRS. 
This is shown in both highly blue and red-shifted ends at $\rm \sim 0.0-2.3 ~km~s^{-1}$ and  $\rm \sim10.2 - 14.0~km~s^{-1}$, respectively.
We believe that this may be a part of of weak bipolar jets coming out from L328-IRS,  projected to some degree to the line of sight.

The other noticeable feature is a blue-shifted X shape structure in both northern and southern regions which is actually a typical feature of the bipolar outflow from a protostellar object. This structure appears  at the velocity of $\rm \sim 2.4 ~km~s^{-1}$ and is seen up to $\rm \sim 6.2~ km~s^{-1}$ .
Then this is getting mixed with other structures, probably some parts of conic outflow shapes including an arc structure in south region  in the range of  $\rm \sim  4.9 ~km~s^{-1}$ to $\rm \sim  6.2 ~km~s^{-1}$.  
 
 L328-IRS has a dense gas envelope of a few tens of arc seconds in size traced by dust continuum at 350 $\micron$ (Lee et al. 2009) or high density molecular line tracers such as $\rm N_2H^+$ 1-0 (Lee et al. 2013). 
The velocity range in $\rm N_2H^+$ line profile for this envelope identified by single dish observations is about between $\rm \sim 6.2 ~km~s^{-1}$ and $\rm 7.5 ~km~s^{-1}$ as shown in Figure 2 of Lee et al. (2013).  However, there is no emission between $\rm \sim 6.3 ~km~s^{-1}$ and $\rm 7.2 ~km~s^{-1}$ in our ALMA data, and thus this no emission is thought to be most likely due to the resolved-out of the envelope emission surrounding L328-IRS to be traced at relatively larger scale  than the recoverable size (11$\arcsec$.4) in our observations.
 
The outflow components which are now red-shifted appear again from the velocity of $\rm 7.3~ km~s^{-1}$.
The most striking features in red-shifted outflow are conic and arc shapes in south and north directions, respectively. Southern conic outflow structures are seen between about $\rm 7.3 -8.8~km~s^{-1}$ and 
northern arc structure appears from about $\rm 7.5~km~s^{-1}$ up to $\rm10.2~km~s^{-1}$. 

All these structures in the CO map are probably compressed parts of the bipolar outflow or just edges of the bipolar outflow where the column density becomes larger to the line of sight. 

In Figure 3 we plot these various blue- and red-shifted outflow components  seen as X, arc, and conic shapes in a single image with the dust continuum emission in contour to help to understand overall structures of the outflow and the continuum emission. 
This figure now shows how blue- and red-shifted conic outflows in southern lobes are well coincident with each other and how a blue-shifted conic feature  in northern lobe is not well matched with a red-shifted arc shape.
We note that the continuum source is located on the tip of blue- and red-shifted lobes, and  the long axis of the continuum emission (P.A. $\sim 106^{\circ}$.5) is almost perpendicular to the axis of the outflows (P.A. $\sim 11^{\circ}$ as given in section 4.2). 
This indicates that the continuum emission is likely a circumstellar disk for L328-IRS.

\subsubsection{$\rm ^{13}CO$ emission}
$^{13}$CO  line emission is not as bright as CO, and thus  is  seen only over the dense parts of the outflow close to the center of L328-IRS.
This feature can be seen in its channel maps or a moment 1 map  of $^{13}$CO emission in Figure 4.
Blue-shifted outflows appear  at the  velocity range of $\rm 4.6 - 6.1~km~s^{-1}$ while the red-shifted outflows are seen at the  velocity range of $\rm 7.0 - 8.0~km~s^{-1}$. 
The other velocity components at the velocity of  $\rm 4.6 - 6.1~km~s^{-1}$   and $\rm 8.0 - 9.3~km~s^{-1}$ are distributed over the disk-like continuum structure and thus likely related to rotational motions
of disk-like components. 
Red-shifted outflow component is also seen at $\rm 7.4 - 8.2~km~s^{-1}$ along a direction of major axis of the continuum emission. This is likely a dense part of redshifted CO outflow in conic shape shown near to the disk-like 
continuum emission as shown in Figure 2 and 3. It should be noted, however, that the moment 1 map actually shows some velocity gradient across the major axis of the disk-shape continuum and thus $^{13}$CO emission may be 
tracing some rotational motions in the structure of the disk-like continuum while no such velocity gradient is seen in the CO emission map.  So it seems that $^{13}$CO emission gives a velocity structure for the disk-like continuum, but
at the same time the outflow activity near the disk-like structure makes it complicated to properly analyse the kinematics around the disk-like continuum.
We will explain more about this  in the discussion section.
 
\subsubsection{$\rm C^{18}O$ emission}
 C$^{18}$O emission is known to be  one of the best tracers for investigating the kinematics around the protostellar disks (e.g., Ohashi et al. 2014; Aso et al. 2017) and thus thought to be useful for examining the gas motions 
 around disk-like structure near to L328-IRS.
 Figure 5 plots  the moment 0 map of C$^{18}$O emission with dust continuum emission in a contour of its 5$\sigma$  level ($\sim 0.2$ mJy beam$^{-1}$),
 showing that overall C$^{18}$O emission traces fairly well the dust continuum region and the major axis of the CO emission distribution looks almost parallel to that of  the continuum emission distribution in the NW-SE direction. 
 
The overall kinematics near to L328-IRS can be well examined by the moment 1 map of the C$^{18}$O emission shown in Figure 6. 
This figure clearly displays a velocity gradient through the direction of the elongated C$^{18}$O emission 
which is very coincident with the major axis of dust continuum emission drawn in a single contour of its intensity, while there is no clear velocity gradient along the short axis of the C$^{18}$O emission distribution. 
If the disk is rotating, there should be the largest velocity gradient along the major axis while there is no velocity gradient along the minor axis. 
These velocity features imply that  a dominant kinematics  in the disk-like shape C$^{18}$O emission may be a rotating motion.
Therefore C$^{18}$O emission will be used in further analysis of the rotating kinematics in the disk structure of L328-IRS in next sections.\\ 

However, we should note that there are some complications  which are needed to be considered in interpreting the rotational motions around L328-IRS using C$^{18}$O emission. 
One complication may come from the fact that  the C$^{18}$O emission is not centrally peaked, but more likely flattened to the NW and SE directions, while the continuum emission is centrally peaked. 
We looked at C$^{18}$O spectra at NW and SW parts of the disk, finding that the C$^{18}$O emission is significantly mixed with the other velocity components forbidden from the disk rotation. 
Such components can be also noticed in a position-velocity diagram along the major axis of the disk structure  presented in forthcoming section.
  Therefore this flattened C$^{18}$O distribution to the NW and SE directions is thought to be mainly due to the existence of these forbidden components in those directions  
 (in addition to the rotational components) which may be mostly outflow components coming out from L328-IRS.
These forbidden components can be also seen in the moment 1 map of C$^{18}$O emission  in Figure 6 as the red-shifted ones near the western edge of the disk structure. 
 
The other complication may come from the presence of C$^{18}$O components (mostly in the velocity range of  $\rm 7.1 - 7.3  ~km~s^{-1}$) to the NE direction  that are not seen  in the disk-shape continuum feature. 
Figure 6 indicates that these components may be some parts of the outflow emanating from L328-IRS to the NE direction of the disk. 

The third complication may come from the  fact that  the direction (P.A. $36^{\circ}.5$) toward which there is no velocity gradient is slightly different from that (P.A. $16^{\circ}.5$) of the minor axis of the continuum disk, as shown 
in the moment 1 map of C$^{18}$O emission.
This is probably due to the presence of  other motions such as outflows and/or infalling motions which can  disturb the rotational kinematics in the disk.

Therefore, in interpreting the kinematics of disk-like structure in C$^{18}$O emission, it is important to consider that the C$^{18}$O disk of L328-IRS is not in purely rotational motions, but contaminated with these complex velocity  components which have nothing to do with the kinematics of rotating disk structure.

The rotating and other kinematical features are also shown in C$^{18}$O channel maps in Figure 7. The blue-shifted components approaching to us are clearly seen in the velocity range of $\rm 3.7 - 5.1  ~km~s^{-1}$ while the red-shifted components moving  away from us are seen in the velocity range of $\rm 8.7 - 10.3  ~km~s^{-1}$. 
We note that the faster blue or red-shifted velocity components are at inner disk part and the slower blue or red-shifted velocity components are at outer disk part, indicating that the disk may be in Keplerian rotation.  
However, we also note that some components in the range of $\rm 5.2 - 8.6 ~km~s^{-1}$ are distributed well over the disk-like structure and likely regarded as other kinematical components like outflows and/or 
infalling structures, making an interpretation of disk kinematics to be more complicated. More detailed analysis on the kinematics in the disk-like structure is given in the  discussion section.

\subsubsection{$\rm CO$ isotopologue line profiles toward disk-like continuum structure of L328-IRS}
A spectral line profile is another useful tool in diagnosing various kinematical features toward a protostellar object and its gas envelope. 
CO isotopologue lines are particularly useful to identify motions related to outflows, infall, and rotation toward protostars and their surrounding envelopes. 
Figure 8 plots CO isotopologue line profiles toward the most significant part of $~1\arcsec.0 \times 0\arcsec.6$ of L328-IRS  where the various gaseous motions such as  the outflows,
possible disk rotation, and infalling motions from its envelope may exist. \\

The CO 2-1 profile indeed shows a variety of interesting features. The line profile spans over wide velocity range almost between 0 and $\rm \sim 14~km~s^{-1}$ with absorption features  in this range.  By comparing the channel maps of CO and  C$^{18}$O emission shown in Figure 2 and 7 with this CO profile, it is thought that most of wide wing parts of the profile are the outflow components from L328-IRS,  and the main bright profiles  between about $\rm 3.0 - 6.2 ~km~s^{-1}$ and $\rm 7.5 - 10.0 ~km~s^{-1}$ are mostly the rotating disk component of L328-IRS. In addition, CO profile shows a prominent absorption feature in the velocity range of $\rm \sim 6.2 - 7.5 ~km~s^{-1}$. As mentioned in previous section, this absorption feature is due to the resolved-out phenomenon of the envelope emission surrounding L328-IRS 
at our ALMA observation which has very similar velocity span to the velocity range of the absorption profile. In the CO profile there is an even deeper absorption feature having negative intensity at $\rm \sim 7.4~km~s^{-1}$, so called an inverse P Cygni profile.
 This is possibly due to an absorption of the disk continuum emission or resolved-out envelope emission  by the red-shifted ``cool'' foreground envelope gas which is infalling toward L328-IRS. We suggest that 
 the latter cause is more likely because such absorption profile at similar velocity is also seen at various directions as well as the direction of the central disk-like structure (Figure 9). It is noted that the absorption peak velocity of this profile is coincident with that ($\rm \sim7.2~km~s^{-1}$) of  the red-shifted profiles (which may be red-shifted infalling gas components) seen in infall asymmetric spectra of CO 3-2 and HCN 1-0 in the single dish observations (see the Figure 2 of Lee et al. 2013).

If the systemic velocity is assumed to be 6.6 $\rm km~s^{-1}$ as derived in next section, then infall speed  will be roughly about 0.8 $\rm km~s^{-1}$.
 This inverse P Cygni profile is also seen in $^{13}$CO and C$^{18}$O lines, but with somewhat different extents. In $^{13}$CO  the absorbed depth is shallower than that of CO line and its absorption peaks at $\rm 7.2 ~km~s^{-1}$ which is slightly less red-shifted.      
In C$^{18}$O  this absorption feature is weak, but enough to be recognizable, with its smallest red-shift among CO isotopologue lines.
We will further discuss this profile and its implication in the next section.\\

The $\rm ^{13}CO$ 2-1 profile shows a similar feature to that of the main isotopologue line CO 2-1, but with some differences from the CO line profile.
One main difference is that $\rm ^{13}CO$ 2-1 has narrower line wings than CO line, covering the velocity range between $\rm \sim 2.2 - 11.0~km~s^{-1}$ where 
emitting components are between  $\rm \sim 2.2 - 6.4~km~s^{-1}$ and  $\rm \sim 7.4 - 11.0~km~s^{-1}$. This velocity coverage is still wide enough to see outflow emission at the farthermost velocity range 
from the systemic velocity and  $\rm ^{13}CO$ emission near the systemic velocity is bright enough to investigate some kinematics of innermost possible rotational motions of the disk-like emission structure. 
There is a feature of the inverse P Cygni profile which seems to be deep enough to infer infalling motions of gaseous material.  
The velocity  range of the absorption feature due to the resolved-out foreground envelope emission is slightly smaller than that of the CO line.

The overall feature of the most rare isotoplogue C$\rm ^{18}O$ 2-1 profile appears to trace a similar velocity range to that by  $\rm ^{13}CO$ 2-1, but the details are somewhat  different from $\rm ^{13}CO$ 2-1 line profile.
For example, the velocity range of C$\rm ^{18}O$ emission is  slightly narrower than $\rm ^{13}CO$ emission, covering  $\rm \sim 2.2 - 6.4~km~s^{-1}$ and  $\rm \sim 7.1 - 10.5 ~km~s^{-1}$. 
Thus C$\rm ^{18}O$  emission may hardly trace outflow components at high velocity range than the  $\rm ^{13}CO$ emission, and be less contaminated with other velocity components such as the outflows. 
In fact C$\rm ^{18}O$ 2-1 line seems to well trace the central disk-like component without any significant contamination by other components such as the outflow.
But it is weaker than $\rm ^{13}CO$  and thus may be less favorable in this regard in analysing the kinematics for the disk component.
In comparison $\rm ^{13}CO$ shows stronger disk emission and thus can be better used for investigating its kinematics than the C$\rm ^{18}O$ emission if other contaminating velocity components are properly extracted, when the C$\rm ^{18}O$ emission is too weak.

\section{Discussion}
    
\subsection{Rotational kinematics in the disk and its implication}

In the previous section we examined  the kinematics in the disk-shaped structure seen in dust continuum emission by using the moment 1 and the velocity channel maps of C$^{18}$O emission, suggesting 
that the disk may be possibly in Keplerian rotation, with some  contamination of other kinematics such as infall and outflow motions.
In this section we  discuss whether the rotational motions in the disk-shaped structure follow a Keplerian kinematics or not  and how the other motions can affect the rotating motions in the disk-shaped structure.

For this discussion we constructed a position-velocity (PV) diagram (Figure 10) along the major axis (P.A. $106^{\circ}.5$) of the continuum disk in order to derive the rotational velocity of the disk emission
as a function of the radius from L328-IRS.
The linear distance of the emitting position from L328-IRS was estimated by assuming the distance of L328-IRS from the Sun to be $\sim 217$ pc (Maheswar et al. 2011). 
In order to get the rotational velocity at each radius we made a cut profile along the velocity axis at each position in the PV diagram which became a kind of an intensity spectrum as a function of the LSR velocity at a given radius.
We fitted  each spectrum with a Gaussian function to obtain the velocity at its peak intensity  at each position. As shown in the moment 1 map of C$^{18}$O, there are other kinematical contaminants in a  disk rotation such as gas outflows or infalling motions. Some of those velocity components  can be seen as  forbidden in the rotational kinematics in the PV diagram. 
These are the components at the first and third quadrants in the left panels of Figure 10 and 11 and neglected in our analysis of the disk rotation.
 This way pure rotation components only could be extracted for the study of the rotational kinematics of the disk structure. 

The Gaussian fit velocity was subtracted from the systemic velocity of the disk in order to obtain the projected rotational velocity of the disk.
In this procedure the systemic velocity of the disk was determined to minimise  the total summation of the differences between two Gaussian fit velocities in blue-shifted and red-shifted parts at the same radii of the disk from L328-IRS. Our obtained systemic velocity is 6.61 $\rm km~s^{-1}$ which is found to be very close to the systemic velocity value ($\rm \sim 6.7~km~s^{-1}$) of the envelope of L328-IRS obtained by a single dish observation (Lee et al. 2013).
The rotation velocity was  derived from the projected rotational velocity assuming that  the disk is inclined as $25^{\circ}.2$ to the far direction of the sky plane which is determined  from $i= \arcsin({b\over a})$ where a ($=40\arcsec$) and b ($=17\arcsec$) are semi-major and minor axes of the deconvolved disk.  
In estimating  the inclination angle of the rotating disk of L328-IRS  we assume that  the disk is in circular shape when it is viewed face-on and its observed shape is the one projected on the sky.

The constructed rotational velocity diagram for  gas motions along the major axis of the rotating disk is given at the right hand side of Figure 10. There, the rotational velocities (that are observationally derived) are plotted as a function of radius along the major axis of the disk-like structure  with those for theoretical Keplerian motions in the disk by the central point sources of 0.05 to 0.4 $\msun$.  A puzzling  feature in Figure 10 is that the disk-like structure shows 
increasing velocities from 130 AU  to 60 AU radius while it has decreasing velocities from 60 AU to central region of the disk. We found our data between 130 AU and 60 AU are the best fitted with the Keplerian motions in the disk by a stellar object of $\sim0.30 ~\msun$.
However, in contrast, the rotational velocity  is not fitted with Keplerian motions at all  toward the center of disk from $\sim 60$ AU radius. Its decreasing pattern within 60 AU is close to that of solid rotation although it is very unlikely that there is the solid rotation in the inner part of the gaseous disk. 

We made a similar analysis using $^{13}$CO line data.  $^{13}$CO line emission is brighter than C$^{18}$O emission and thus may be more useful in the kinematical analysis of disk emission faint in C$^{18}$O line. 
However, at the same time it may suffer more from the contamination by outflow motions. We made its PV diagram along the major axis of the disk emission and the rotation velocity versus disk radius diagram using the same procedures 
as described above for C$^{18}$O data (Figure 11). 
We found very similar results seen in diagrams for C$^{18}$O line data in Figure 10: Keplerian motions from 140 AU  to 60 AU radius by a central object of $\sim 0.27\msun$ and decelerated motions from 60 AU radius to central region of the disk.

At this moment it is hard to explain how outer parts are moving like the Keplerian motions while inner parts  are not.
One thing making the interpretation of  the  kinematics in inner disk region difficult is that those parts are being affected by outflows and/or infalling motions. 
We did not include the gas components forbidden by the disk rotating motions  as noted in this section, but it may be possible that our data are still including the components by the outflow or infalling motions at the disk parts very close to the central object 
which seem to somehow follow disk rotating direction, but having a large departure from the rotating motions. 
In fact Machida et al. (2009) showed from their MHD simulation for the formation of BD in a compact cloud core that these complicated motions can exist within a few tens AU radius from the BD and thus can likely affect the interpretation of disk rotation motions
if the observing lines do not trace dense disk parts only.  

There may be a possibility that multiple sources inner disk region not resolved in the present observations can exist and thus make the kinematics within a few tens AU radius more complicated. 

Further study in future  with more appropriate line observations in better spatial resolution which can better trace rotational motions of innermost parts without any contamination by other motions may be extremely useful.
For example, SO and $\rm H_2CO$ lines are found to be centrally enhanced in the protostellar disk and thus may be potentially good to trace inner kinematics in the disk (Ohashi et al. 2014,  Sakai et al. 2014).
With this kind of tracers  in higher spatial resolution better clarification for the gas kinematics in the disk structure of L328-IRS may be given in near future.

\subsection{Overall system of L328-IRS}   

All astronomical  features are observed in projection on the sky to the line of sight. Thus this always becomes an obstacle in making a right interpretation on the physical characteristics of observed objects.
However, this difficulty can be lessened or even overcome if the system has a symmetric shape and thus its 3D feature can be reasonably inferred. 

Here we apply this idea for L328-IRS system and discuss how this system having the infalling envelope, the outflows, and the disk,  looks in  three dimensional space in order to attempt to draw a better  inference on its whole structure and the related physical features.
The key benefit of this process is to enable us to reliably infer the inclination angle of the outflow cavity which is hardly known from single dish observations, and thus  help to significantly reduce the main uncertain factor of the mass accretion rate.

The constructed 3D system of L328-IRS should be able to explain most of the observational characteristic structures shown in Figure 4 and 5, especially 
various outflow features  such as X shape cavities,  blue and red-shifted arc structures seen in South and North with respect to the disk structure, and inverse P-Cygni profiles at various directions as well as the direction toward L328-IRS.

For this purpose we devised a simple toy model for L328-IRS. The model has a shape of the trapezoidal rotating body as shown in Figure 12-(a) to mimics a simple shape of outflow cavity. It consists of two cavities having a uniform thickness 
of $0\arcsec.45$ which is assigned from the thickness of X-shape and arc structures seen in the CO map.  The cavity structures have a spatial size of $15\arcsec \times 15\arcsec$ consisting of $300 \times 300$ pixels$^2$ which is the same spatial size containing the same pixels$^2$ as our observing data. All the outflow emission is assumed to come from this thickened cavity structure only. It is also assumed in our simple toy model that all pixels in charge of the outflow structures emit the same intensity, there is no radiative transfer effect in the model system, and thus the line-of-sight intensity will be only the sum of emission from the line-of-sight pixels.

In this simple model an exact velocity information could not be assigned for the model outflow structures. Instead, we could make a simple inclusion of  at least blue- or red-shifted velocity information in the model outflow, by assigning a blue-shifted velocity for the  components at the nearer side from an observer with respect to the sky plane and a red-shifted velocity for the other components  at the farther side from the observer with respect to the sky plane as indicated in Figure 12-(a).  

Our observed data to compare with the model are composed of a  blue-shifted intensity map integrated between $\rm 2.24 - 6.24~km~s^{-1}$ and a  red-shifted intensity map integrated between $\rm 7.78 - 10.4~km~s^{-1}$.

The simplest observational constraint we can make in the beginning for a search of the best model would be the position angle of the axis of the overall outflow structure which turned out to be  $\sim 11 ^{\circ}$ from its observed shape. 

The next  parameters we can observationally constrain are the opening angle of the outflow cavity and the inclination angle of the outflow axis. Difficulty in constraining two parameters is that they are dependent on each other in a way that 
the projected value of the opening angle of the outflow cavity is highly affected by how the outflow axis is inclined.  Thus we examined the models with various inclination angles and opening angles in their wide parameters space  to find the best model which can reproduce the observed intensity distribution of L328-IRS system.

The parameter space for the opening angle was examined between $75^{\circ} - 105^{\circ}$ around  the likely value  ($90^{\circ}$) guessed from the observed X shape cavities. On the other hand,  the best value of the inclination angle was searched between $50^{\circ} - 80^{\circ}$  because the most possible value would be around at $65^{\circ}$ which is guessed from the disk inclination. 

The aspect ratio of the disk emission itself gives a degeneracy in inferring  the inclination angle of the disk. In other words the inclination angle can be either  $65^{\circ}$ or $115^{\circ}$. 
However, the blue and red-shifted arc structures shown in Figures 4 and 5 can be possibly made only if the compressed or dense material exists in the red-shifted northern cone and the blue-shifted southern cone whose inclination (or the normal vector of the disk) is close to $65^{\circ}$. Therefore we attempted to search for the best parameter for the inclination angle  around $65^{\circ}$.

In this procedure one of important observed features that we need to reproduce is the arc-structures seen in northern and southern cones of the outflow cavity. We assumed that there exists the dense material in arc-shape in the cones that  is distributed in a separate structure  in the northern and southern cones at the same distance from the center of the disk. The nearest distance of the arc structures from the center of the disk is given $\sim 1\arcsec.5$ as observed.

In searching for the best parameter set for the opening angle of the outflow cavity and its inclination angle we multiplied intensities from the same pixels between each model and the observed data, and summed them. This sum of the multiplied intensities between each model and the observed data  is found to be a useful measure for examining a reproducibility of observing data by a model. The model giving the highest value will be the one best reproducing the observation. 

In the beginning we calculated the sum of the multiplied intensities between the observed data and  49 models at  $5^{\circ}$ interval  in the two searched parameter ranges to find a peak value of the sum of the multiplied intensities in the ranges of the opening angles of $85-100^{\circ}$ and the  inclination angles of $65-75^{\circ}$. Then we re-calculated the sum of multiplied intensities between observed data and  256 models in a finer interval of $1^{\circ}$ in these parameter ranges, finally finding that the peak value of the sum is produced at a model which has an opening angle of  $92^{\circ}$ and an inclination angle of $66^{\circ}$.  Figure 12 (b) shows this result in a relative ratio of the sum value of each model with respect to the sum value of the the best model, indicating that two parameters are very well constrained. This figure displays  that  variation of  $3-5^{\circ}$ for the opening angle and the inclination angle from the angle values of the best model outflow cavity will reduce $\sim 6\%$ of the sum value at the peak.

The  disk itself with a position angle of $106^{\circ}.5$ as measured in the previous section is rotating such that  the eastern part is moving away from us while the western part of the disk is moving toward us.
 Therefore we are probably seeing the bottom part of the disk which is in counterclockwise rotation. We also notice that the disk is located on the bases of the outflow not in a symmetric center of the outflow bases 
 such that western parts of the bases are longer than the eastern bases as mentioned above, implying that the eastern side of the disk  are farther  from us than the western part.  

One promising kinematical feature shown in L328-IRS system is infalling motions in gaseous envelope. 
In the previous section we showed inverse P Cygni profiles in the CO lines implying existence of infalling motions of envelope gas material. 
These characteristic features are seen nearly all over the gas envelope of L328-IRS except for 
the regions where outflow activities are strong, with some variability of the negative intensities from region to region as shown in Figure 3. This indicates that infalling motions of gaseous material traced in CO 2-1 are occurring nearly all over the envelope of L328-IRS.
Although this  is not dealt with in our simple toy model, we marked the infalling features in the model system of Figure 12-(a).
The overall schematic 3D view of L328-IRS system inferred from this work is depicted in Figure 12-(a).

\subsection{ALMA estimation of mass accretion rate and accretion luminosity in L328-IRS, and its implication}
A mass accretion rate on a protostellar system is one of the most essential physical quantities to understand what the system would be  and how it will evolve. Lee et al. (2013) have derived this rate for L328-IRS by using single dish data, concluding that  L328-IRS has a very low mass accretion rate ($\sim 3.6\times 10^{-7} \msun~yr^{-1}$ ) which is  an order of magnitude less than the canonical value for a protostar (Shu et al. 1987; Dunham et al. 2006). 

Here we re-estimate the mass accretion rate of L328-IRS with ALMA data. ALMA observations did not cover the whole area of the outflow, but they have an advantage that the ALMA  filters out large scale envelope components 
and thus can preferentially trace compact outflow blobs.
Therefore the ALMA data may give more accurate estimation of physical quantities of the outflow with the least contamination by the envelope gas component.
The mass accretion rate is estimated with the following equation formulated from the momentum conservation in a protostellar system under a process of the transformation of gravitational energy  by the mass accretion to
wind and/or jet energies (e.g., Bontemps et al. 1996);
$$
\dot M_{acc} = {1\over f_{ent}}{\dot M_{acc} \over \dot M_{W}  } {1\over V_{W}} F_{outflow},
\eqno (2)
$$
where $f_{ent}$ is an entrainment efficiency ($<1$),  ${\dot M_W}$ is a wind/jet mass loss rate,  $V_{W}$ is a jet/wind velocity, and 
$\rm F_{outflow}$ is an outflow force. In this calculation we adopted the same values for $f_{ent}$(=0.25),  ${\dot M_{W} \over \dot M_{acc}}=0.1$, and
the jet/wind velocity $\rm V_{W}\sim150~km~s^{-1}$ as those adopted by Lee et al. (2013) for the comparison.  
 Here the outflow force $\rm F_{outflow}$ is given as 
$$ F_{outflow}={\sum_i \sum_j P_{i,j} \over t_i}  {sin\theta \over cos^2\theta},
 \eqno (3) 
 $$
where $\rm P_{i,j}$ is the outflow momentum
(= $\sum_i \sum_j m_{i,j} (v_{i,j}-v_{sys})$ where $i$ and $j$ are indexes for the summation over the positional pixels and the velocity channels over the outflow, respectively.). Here 
$t_i$ is a dynamical time of an outflow component at i$^{th}$ pixel, the time over which an outflow blob at i$^{th}$ pixel travels the distance ($d_i$) from L328-IRS to its position,
given by 
$$t_i = {d_{i} \over |{v_{i,j}-v_{sys}|}}{cos~\theta \over sin~\theta},
\eqno (4) 
$$
 where $v_{i,j}$ is a LSR velocity of an outflow assigned by j$^{th}$ velocity channel at i$^{th}$ position pixel, 
$v_{sys}$ is a systemic velocity of L328-IRS system derived as  6.61 $\rm km~s^{-1}$ in the previous section, and $\theta$ is an inclination angle of the outflow to the line-of-sight.
The mass $m_{i,j}$ of the outflow blob with j$^{th}$ velocity channel at  i$^{th}$ position pixel can be obtained by the column density multiplied by a pixel area of each position
as follows;
$$
m_{i,j} =  \mu_{H_2} m_H N_{i,j} S_i ,
\eqno (5)
$$
where $\mu_{H_2}$ is the mean molecular weight per hydrogen molecule ($\sim 2.8$ for gas of a cosmic abundance of  71\%
hydrogen, 27\% helium, and 2\% metals by mass; Cox 2000),  $m_H$ is a mass of a hydrogen atom,  $N_{i,j}$ is the column density of
the outflow blob at  i$^{th}$ position pixel and j$^{th}$ velocity channel, and $S_i$ is the $i^{th}$ pixel area.
Here $N_{i,j}$ can be derived using CO 2-1 line observations by a following equation given for a transition of CO from a lower state J to an upper state J+1 with assumptions of   Local Thermodynamic Equilibrium (LTE) condition and a thin optical depth toward the  outflow blobs in CO 2-1 profile  (e.g., Dunham et al. 2014); 
 $$
\rm N_{i,j}= X_{CO} {3 k \over 8 \pi^{3} \nu \mu^{2}} {(2J+1)\over (J+1)}  {Q(T)\over g_j} e^{E_{J+1}\over kT} T_{12} (i,j)~\delta v
~ \rm [cm^{-2}], \eqno (6)
$$
where $\rm X_{CO}$ is an abundance of CO relative to $\rm H_2$ (assumed to be $10^{-4}$, Frerking, Langer, \& Wilson 1982), k is the Boltzman constant, $\nu$ is the frequency for CO 2-1 line (230.5380 GHz), $\mu$ is 
the dipole moment of CO (0.1098 debye), Q(T) is a partition function given by $Q(T)=\Sigma_{J=0}^{\infty} g_J e^{E_J/kT}$ at a temperature T, $g_J$ is a degeneracy given by 2J+1, $\rm T_{12} (i,j)$ is an intensity in K in CO 2-1 line, and $\delta v$ is a velocity channel width.
 
For the accurate calculation of the outflow force (or mass accretion rate) we attempted to reduce any possible uncertainties of the quantities given or implied in the equation (6).
For example, the equation (6) is  applicable only when CO 2-1 line (even in its wing parts) is optically thin. However, CO 2-1 line profiles often get saturated (e.g., Arce \& Goodman 2001; Curtis et al. 2010) and thus we need to make a proper correction on the high optical depth effect in CO 2-1 line. 
For this correction $\rm C^{18}O$ 2-1 line data can be useful  as  $\rm C^{18}O$ line has the least optical depth among the isotopologue lines which we observed with. However we  used  $\rm ^{13}CO$ 2-1 emission  because $\rm ^{13}CO$ 2-1 is  detected fairly well over the outflow area around L328-IRS while $\rm C^{18}O$ 2-1 is detected only over the disk area of L328-IRS.
$\rm T_{12} (i,j)$ can be opacity-corrected from the following equation under the assumption that the $\rm ^{12}CO$ and $\rm ^{13}CO$ are at LTE with the same excitation temperature and 
the $\rm ^{13}CO$ 2-1 is optically thin; 
$$
\rm T_{12} (i,j) = T_{13} (i,j)  {{1-e^{-\tau_{12}(i,j)}}\over \tau_{12}(i,j)} {[^{12}CO]\over [^{13}CO]} ,
\eqno (7)
$$
where $\rm T_{13} (i,j)$ is an intensity in K for $\rm ^{13}CO$ 2-1 line, $\tau_{12}(i,j)$ is an optical depth of CO 2-1 line at  i$^{th}$ position pixel and j$^{th}$ velocity channel, and  $\rm {[^{12}CO]\over [^{13}CO]}$ is the abundance ratio which is assumed to be 62 (Langer \& Penzias 1993). 
In estimation of physical quantities we considered the CO spectra brighter than $5\sigma$. Our derived optical depths for CO were between  11 and 62.0 with mode values of 26.9 and 33.6 for the blue and red components, respectively, 
meaning that CO 2-1 line emission is mostly optically thick. 

This high optical depth in CO 2-1 line leads us to estimate excitation temperatures for the outflow blobs which can also help to reduce the uncertainty in our calculation of the mass accretion rate. We estimated excitation temperatures for the outflow blobs to be about 9.9 to 38.2 K.
We used these excitation temperatures and optical depths for the regions where $\rm ^{13}CO$ 2-1 is detected  to derive the column density in better accuracy.
However, for the regions where $\rm ^{13}CO$ 2-1 is not detected, we simply assume that CO 2-1 line would be optically thin and the excitation temperature would be  as cold as $\sim 9.9$ K , and calculate the column density
by the equation (6). Choosing the other value for the excitation temperature is found to be not sensitive of the final result for the mass accretion 
rate because the outflow force at the central region of L328-IRS where the excitation temperature is properly determined is dominantly high (two order of  magnitude higher  than any other regions where the excitation temperature can not be determined). 
For example, our estimation with the excitation temperature of 38.2 K for the regions where $\rm ^{13}CO$ 2-1 is not detected  gives almost the same total outflow force and mass accretion rate as the results with the excitation temperature of 9.9 K.  

Our analysis of L328-IRS images with the 3D toy model also helps to make a better determination of the mass accretion rate because some of parameters for the outflow structures in 3D are well constrained and thus can be used in the calculation of the mass accretion rate with much less uncertainties. 
For example,  an inclination angle of the outflow is usually poorly known. However, our analysis gives a well constrained value of $\sim66^{\circ}$ as explained in the previous section, reducing an uncertainty in the estimation of the mass accretion rate. 
 
In this way we estimate the outflow forces of all the outflow blobs, from which a total mass accretion rate is derived to be $\sim 8.9\times 10^{-7} \msun~yr^{-1}$.
Our previous value (with inclination angle of $\sim57^{\circ}.3$) from the single dish observations is $\sim 3.6\times 10^{-7} \msun~yr^{-1}$ which would correspond to a value of $\sim 6.9\times 10^{-7} \msun~yr^{-1}$ at the outflow inclination angle of $\sim66^{\circ}$,
indicating that our new estimation of the mass accretion rate with ALMA data is consistent with the value obtained from the single dish observations.

With this mass accretion rate and the mass $\rm M_{*}$ of the central source, an accretion luminosity of L328-IRS can be derived by using the equation ${G M_{*} \dot M_{acc}}  \over {2R}$,
if this luminosity is mostly due to accretion from the disk to the central object L328-IRS (e.g., Baraffe et al. 2009),
where G is the gravitational constant and R is the radius of the central object.
If the disk rotation is purely in Keplerian motions by the mass of L328-IRS of $\sim 0.3~\msun$, its accretion luminosity is estimated to be $\sim 1.3~\lsun$ which is more than one order of magnitude brighter than its internal luminosity $\sim 0.05~\lsun$ (Lee et al. 2009) that stands for the present accretion luminosity. 
So far there is no clear evidence yet for the episodic accretion in L328-IRS system and thus the accretion process may have been made likely in a quiescent manner until the present. 
Then the mass of L328-IRS may be over-estimated by more than one order of magnitude. 

If the internal luminosity is due to our estimated accretion rate in a quiescent manner, the mass of L328-IRS to produce its internal luminosity of $\sim 0.05~\lsun$ can be estimated to be $\sim 0.012~\msun$ (rather than $\sim 0.3~\msun$). 
Provided that only a limited amount (about $20\% - 40\%$, Lada et al. 2007; Andre' et al. 2010) of the envelope mass ($\sim 0.1~\msun$) can accrete onto the central object, L328-IRS would have  the final mass of a brown dwarf. 

There is an alternative way to infer the likely mass of L328-IRS, which is to estimate its accreted mass. The accreted mass may be  the first order approximate value of the central mass of L328-IRS because most of the mass  accretion toward the proto-BD may occur 
at the very early stage and then the initially high mass accretion rate can rapidly (and highly) decrease by  
$10^{2} - 10^{6}$  times in a very short time of $10^{4}$ yr since the formation of substellar object (Machida et al. 2009). 
L328-IRS is believed to have its outflow activity at least during the dynamical time scale  of $\rm \sim 6.4\times 10^4~ {cos~\theta \over sin~\theta}  \approx  2.8\times 10^{4} ~yr$ (at the outflow inclination angle of $\sim66^{\circ}$) from Lee et al. (2013) so that most of the central mass may have been already accreted.
Therefore the total accreted mass would be $\sim 0.023~\msun$ in this case with an assumption of $10\%$ extraction of the accreted mass through the outflow (Pelletier \& Pudritz 1992; Bontemps et al. 1996). This value well matches to the mass  that we estimated from the assumption that the  internal luminosity for L328-IRS is mainly due to the accretion process in L328-IRS, and is again in the regime of substellar mass.    
Table 2 summarises our methods to infer the mass of L328-IRS and their related problems. 

\setcounter{table}{1}
\begin{table}[h!]
\centering
\rotate
\caption{Summary of our mass estimation of L328-IRS } \label{tab:tbl2}
\begin{tabular}{ccc}
\tablewidth{0pt}
\hline
\hline
Method	&	Estimated mass & Notes	\\
\cmidrule(r){1-3}
Dynamical mass	&	$\sim 0.3~\msun$ & Inner disk not fully fitted with Keplerian motion	\\
Mass inferred from $\rm L_{int}$	&	$\sim 0.012~\msun$ & Valid only if $\rm L_{int}$ is due to $\dot M_{acc}$	\\
 Mass accreted during dynamical time	&	$\sim 0.023~\msun$ & Valid only if $\dot M_{acc}$ is constant during dynamical time	\\

\hline
\end{tabular}
\end{table}

In conclusion, these rough methods for inferring the central mass of L328-IRS system seem to suggest that our  estimate of the central mass does not exceed the mass of a brown dwarf. 
However, we should emphasize  that L328-IRS can have a stellar mass to produce its accretion luminosity equivalent to the internal luminosity  if it is currently acquiring mass in a much smaller   $\rm \dot M_{acc}$ at least by an order of magnitude than the past.
 Therefore we can not completely rule out a possibility that L328-IRS can have a stellar mass, too.
 We also note that our estimation can be subject to uncertainties.   
Our estimation of accretion rate can be affected by various uncertain parameters including the entrainment efficiency, the ratio of mass accretion rate to a wind/jet loss rate, jet/wind velocity, and other parameters involved in the blob mass estimation.  
Moreover the mass accretion rate may have  varied during the outflow activity of  L328-IRS, and  the accretion rate during the rest of the Class 0 phase is quite uncertain  (e.g., Machida et al. 2009, Machida and Hosokawa 2013)  so that the assumption of our constant accretion rate in the total accreted mass estimation may not be valid.
In any case a dynamically direct estimation of the central mass by looking for the Keplerian motions in the inner disk region  with future ALMA observations in right tracers may be  helpful in making a more concrete identification of L328-IRS.

\section{Summary}
In the paper we present our observing results on the central region of L328-IRS with  1.3 mm continuum at ALMA band 6 and $^{12}$CO, $^{13}$CO, and C$^{18}$O 2-1 lines.
Although L328-IRS has been known to be a proto-brown dwarf candidate because of its small mass accretion rate (a few $10^{-7} \msun~yr^{-1}$ ) and small envelope mass of $\sim 0.09 \msun$, 
its identity as a substellar object has not been confirmed  because its central mass has never been directly determined. 
In this study we attempted to precisely measure the central mass of L328-IRS as well as its  mass accretion rate by using these ALMA observations in order to help to confirm whether L328-IRS is in 
a regime of substellar object.\\
Our results are summarized as follows;  
 
1. We found a disk-like structure in continuum whose deconvolved size is about $0\arcsec.40  \times 0\arcsec.17  $ ($\sim$87 AU  $\times$ 37 AU). This structure is believed to be a rotating disk around L328-IRS in the sense that 
its major axis is almost perpendicular to the outflow axis and its moment 1 maps in  $^{13}$CO and C$^{18}$O 2-1 lines do not show any clear velocity gradient along the minor axis of the disk structure, but  a significant velocity gradient along its major axis. 
We investigated whether the rotational motions follow the Keplerian kinematics, finding that the motions are only partially  fitted with  a Keplerian motion between 130 AU and 60 AU  by a central object of $\sim0.30 ~\msun$,
but not  for the inner region of $\leq 60$ AU radius. With the mass of $\sim 0.3~\msun$, the accretion luminosity of L328-IRS is estimated to be $\sim 1.3~\lsun$ which is much brighter than its known internal luminosity $\sim 0.05~\lsun$, making 
us to believe that the disk is not likely in Keplerian motions and the central mass of L328-IRS can not be kinematically determined from this study yet.\\

2. Our ALMA data provide detailed bipolar outflow structures emanating from L328-IRS, helping us to better define outflow parameters and thus derive the mass accretion rate  in better accuracy.
For this process we constructed a 3D toy model of L328-IRS  consisting of  an outflow cavity of the rotating trapezoidal shape whose spatial size and pixels are given the same as those of our ALMA data. 
All pixels in this simple model were assumed to emit the same intensity without radiative transfer effect in the model system so that the line-of-sight intensity is the sum of emission from the line-of-sight pixels. The red or blue-shifted velocity information is simply given with respect to the plane of the sky in order to compare its projected structures with our key outflow structures seen in CO observations such as the X shape outflows  and the arc-like structures in the cones.

From this comparison our observed features of the outflow were found to be best reproduced if the outflow has the inclination angle of $\sim 66^{\circ}$ and the opening angle of $\sim 92^{\circ}$.
The additional correction for high optical depth for CO outflow and the estimation of the excitation temperature for CO outflow as well as  this precise constraining the inclination angle of the outflow led us to estimate more reliable value of the mass accretion rate which is $\sim 8.9\times 10^{-7} \msun~yr^{-1}$. This new estimation is found to be fairly consistent with the value derived previously  from the single dish observations.\\

3. By using the newly derived mass accretion rate, we inferred the central mass of L328-IRS in  two indirect ways. 
One is to assume that the internal luminosity is due to the current accretion. From the assumption the central mass of L328-IRS is estimated to be $\sim 0.012~\msun$.
The other way is to estimate the mass accreted  during the dynamical time scale  of $\rm \sim  2.8\times 10^{4} ~yr$ assuming a constant accretion rate. This gives a total accreted mass of $\sim 0.023~\msun$. 
Our indirect ways imply that the central mass of L328-IRS is likely in  the regime of substellar mass. 

However, we can not  rule out the other possibility that L328-IRS has a stellar mass if its current mass accretion rate is much smaller than the average value during its outflow activity.    
Moreover, these methods are  subject to various uncertainties, especially regarding the mass accretion rate.  Thus the  identification of L328-IRS as a proto-brown dwarf must be considered tentative. 
Future direct kinematical measurement of the central mass  with more appropriate ALMA observations is needed to clarify the identity of L328-IRS.
 
\acknowledgments 

We thank an anonymous referee for a thorough reading of our manuscript and giving us very useful comments with which our paper was significantly improved. 
This research was supported by Basic Science Research Program through the National Research Foundation of Korea (NRF) funded by the Ministry of Education, Science and Technology (NRF-2016R1A2B4012593).

\clearpage
\begin{figure}
\plotone{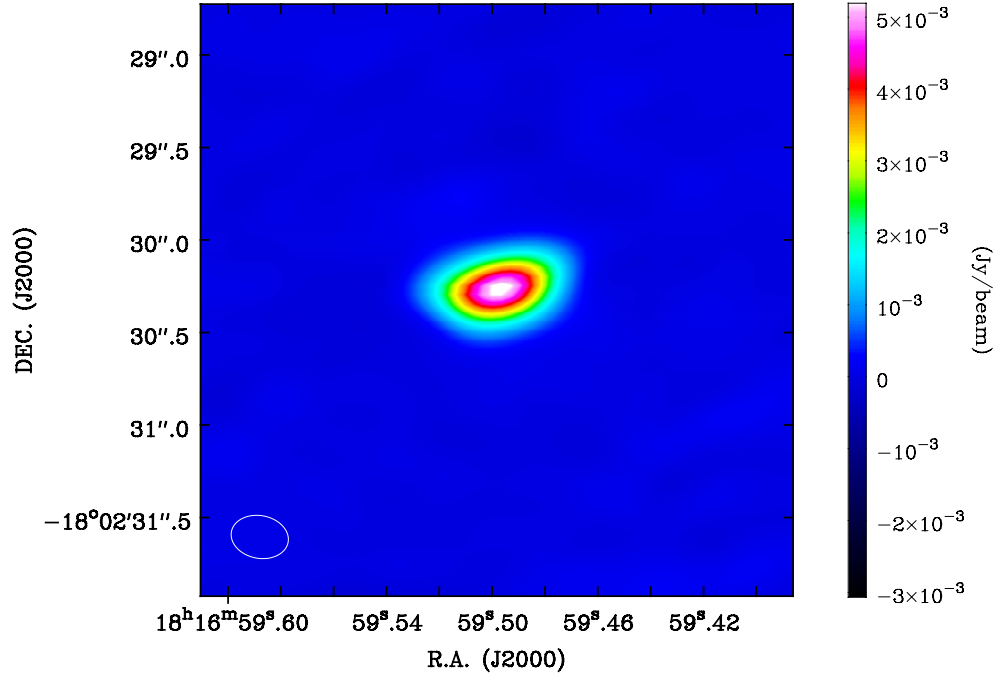}
\caption{
A single compact source showing a ``disk-like'' structure at L328-IRS position in 1.3 mm continuum emission. An open ellipse in the low left corner is the synthesized beam of 0{\arcsec}.31$\times$0{\arcsec}.23(P.A.=79{$^{\circ}.3$}).}
\end{figure}

\clearpage
\begin{figure}
\centering
\includegraphics[height=4.0in,angle=0]{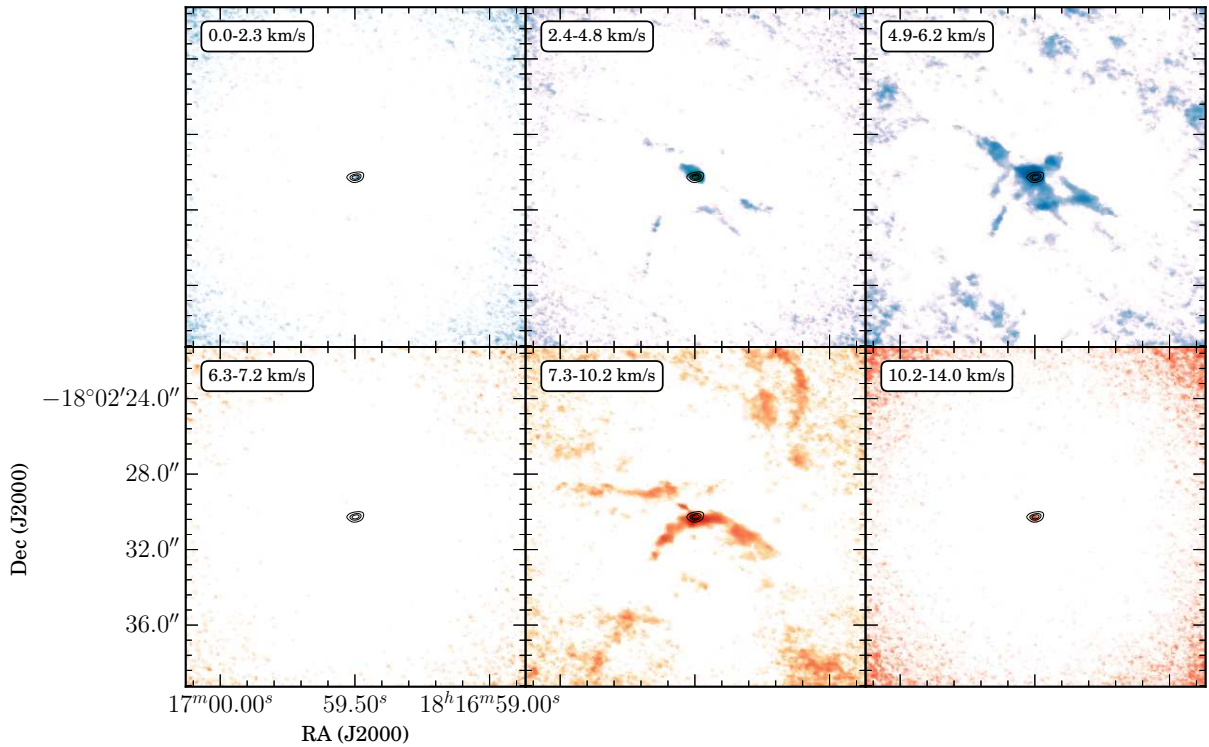}
\caption{
Velocity channel maps of CO emission. In making the map  the emission brighter than $\rm 18 ~mJy~beam^{-1} [0.096~km~s^{-1}]^{-1}$ is considered. Disk shaped continuum is drawn in contour levels, 1.1, 2.2, and 3.3 mJy beam$^{-1}$.}
\end{figure}

\clearpage
\begin{figure}
\centering
\includegraphics[height=6.5in,angle=0]{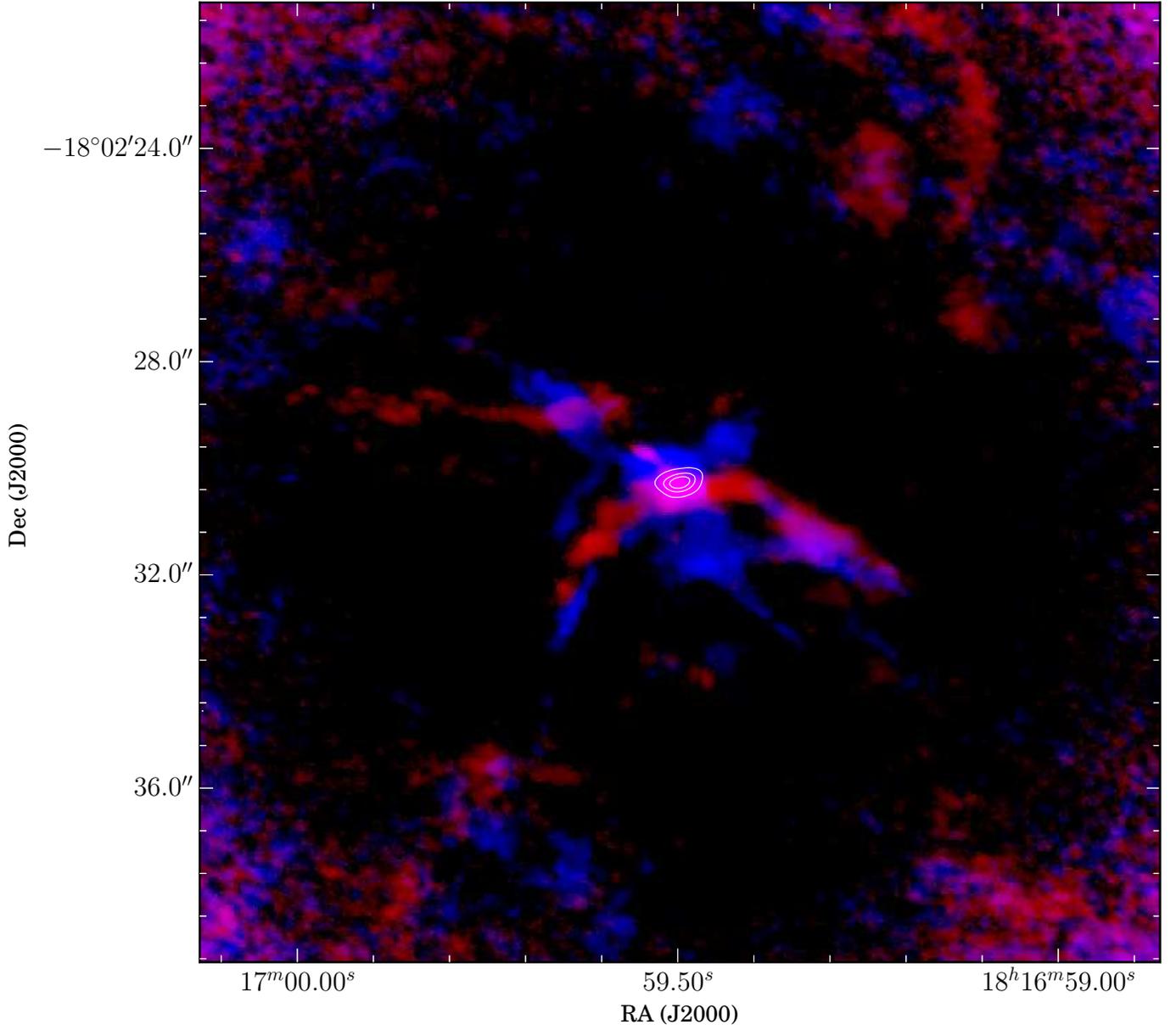}
\caption{
CO intensity map for blue-shifted and red-shifted outflows in colour tones with the dust continuum emission in contours. In making the map the emission brighter than $\rm 15 ~mJy~beam^{-1} [0.096~km~s^{-1}]^{-1}$ is considered.
Contour levels for the dust continuum are the same as the ones for Figure 2.}
\end{figure}

\clearpage
\begin{figure}
\centering
\includegraphics[width=6.00in]{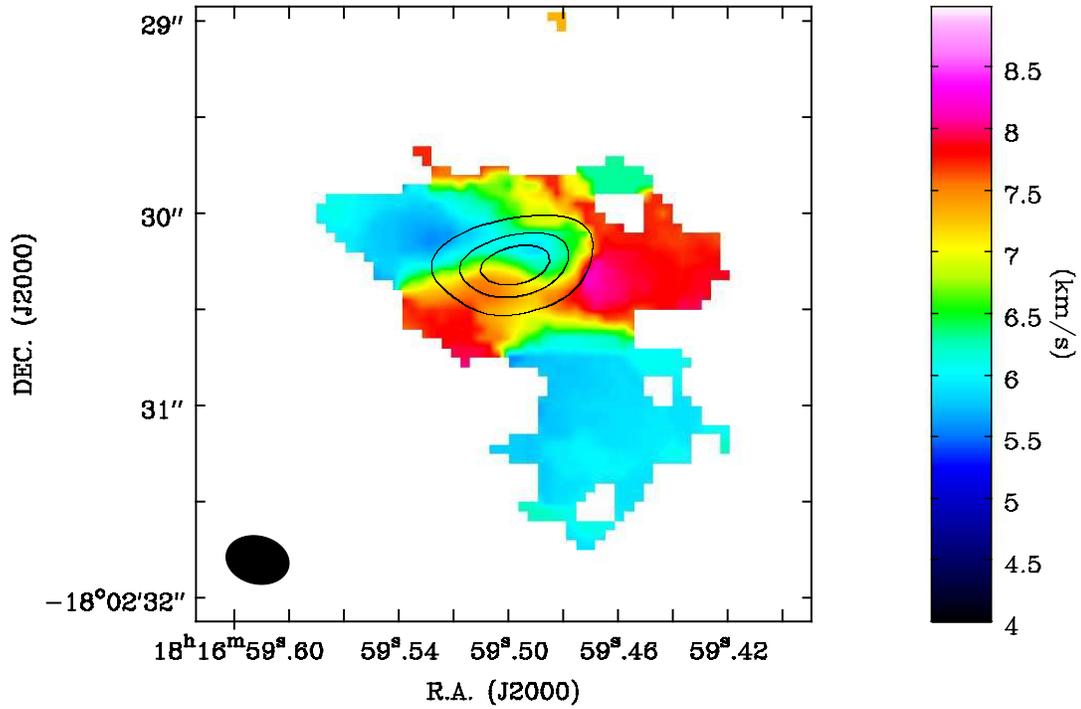}
\caption{
Moment 1 map  of $^{13}$CO emission.  The map shows the emission brighter than $\rm 2.2 \times 10^{-2} mJy~beam^{-1} [0.096~km~s^{-1}]^{-1}$ which is about $\sim 5\sigma$ level of $^{13}$CO emission. 
The dust continuum  is overlaid  in the same continuum levels as the ones in Figure 2.}
\end{figure}

\clearpage
\begin{figure}
\centering
\includegraphics[height=5.0in,angle=0]{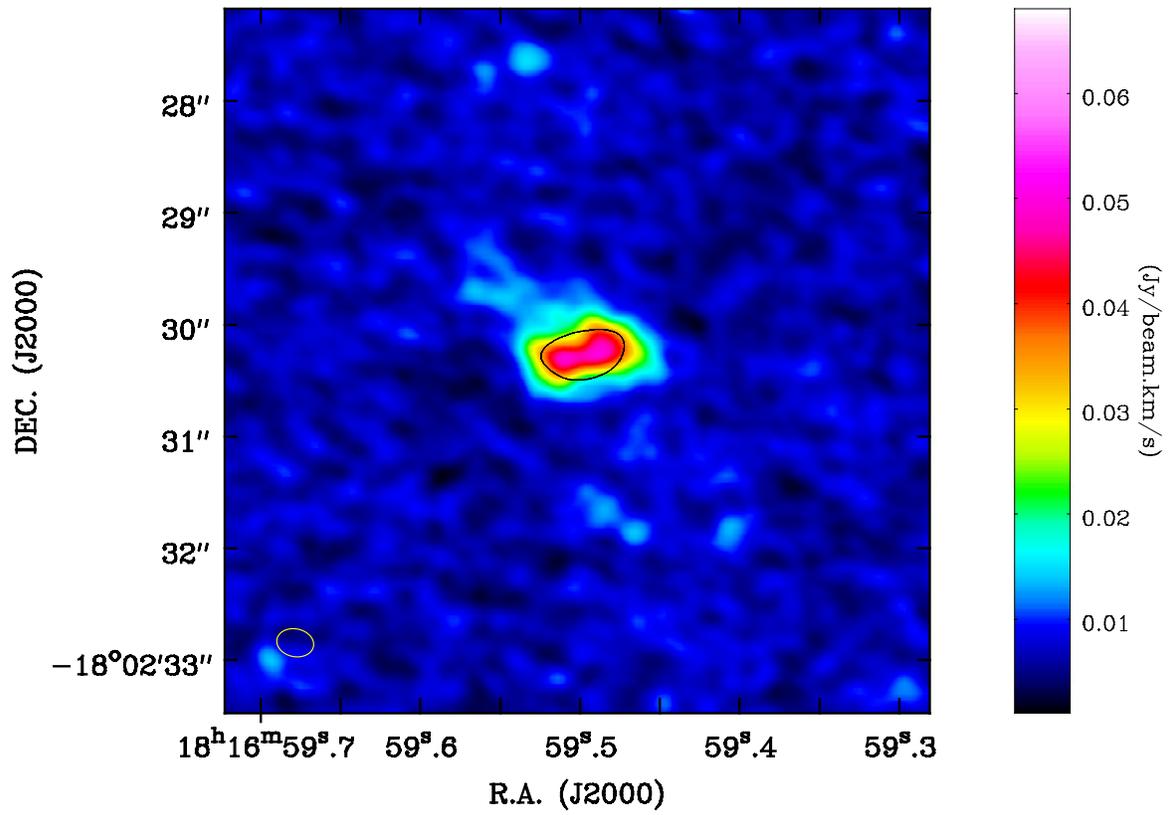}
\caption{
Moment 0 map of C$^{18}$O emission.  Dust continuum emission is overlaid in a contour of $\sim 0.2$ mJy beam$^{-1}$ which is  a 5$\sigma$  level of the intensity. In the map 1 $\sigma$ level of C$^{18}$O emission 
 is about $\rm 5~mJy~beam^{-1} [km~s^{-1}]^{-1}$. }
\end{figure}

\clearpage
\begin{figure}
\centering
\includegraphics[height=5.0in]{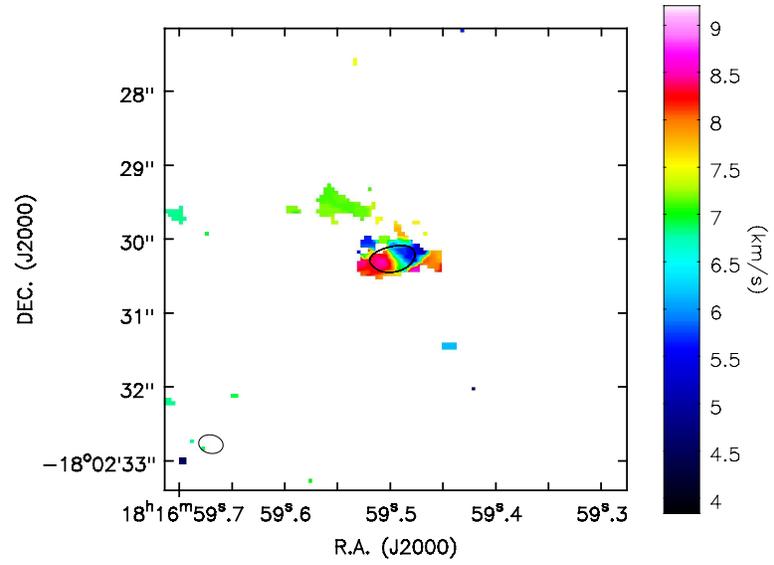}
\caption{
Moment 1 map of C$^{18}$O emission.  The dust continuum  emission is overlaid  in a contour of  its  5$\sigma$  level  ($\sim 0.2$ mJy beam$^{-1}$).}
\end{figure}

\clearpage
\begin{figure}
\centering
\includegraphics[width=7.0in,angle=0]{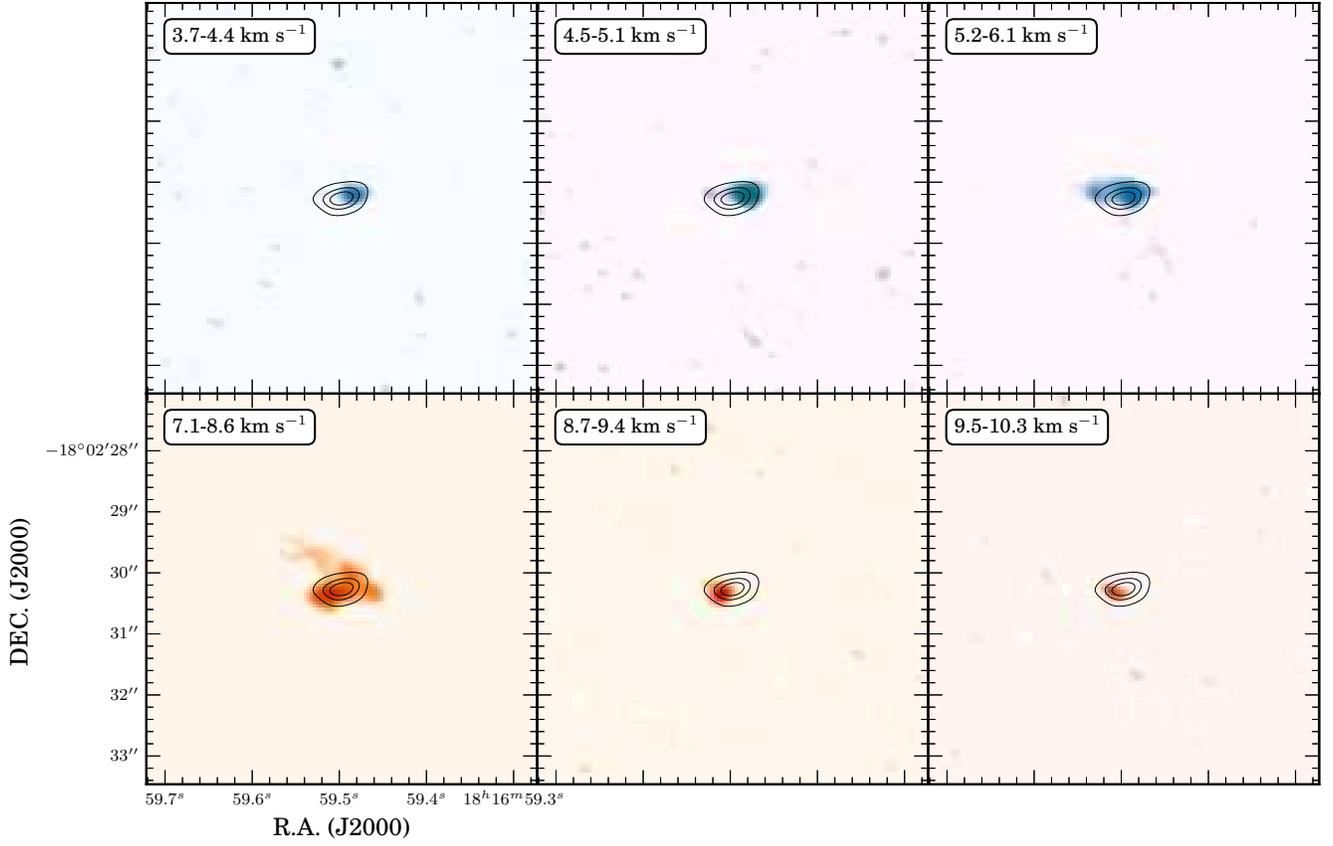}
\caption{
Velocity channel maps of C$^{18}$O emission.  The dust continuum  is overlaid  in contours of the same levels as the ones for Figure 2. }
\end{figure}

\clearpage
\begin{figure}
\centering
\includegraphics[height=6in,angle=0]{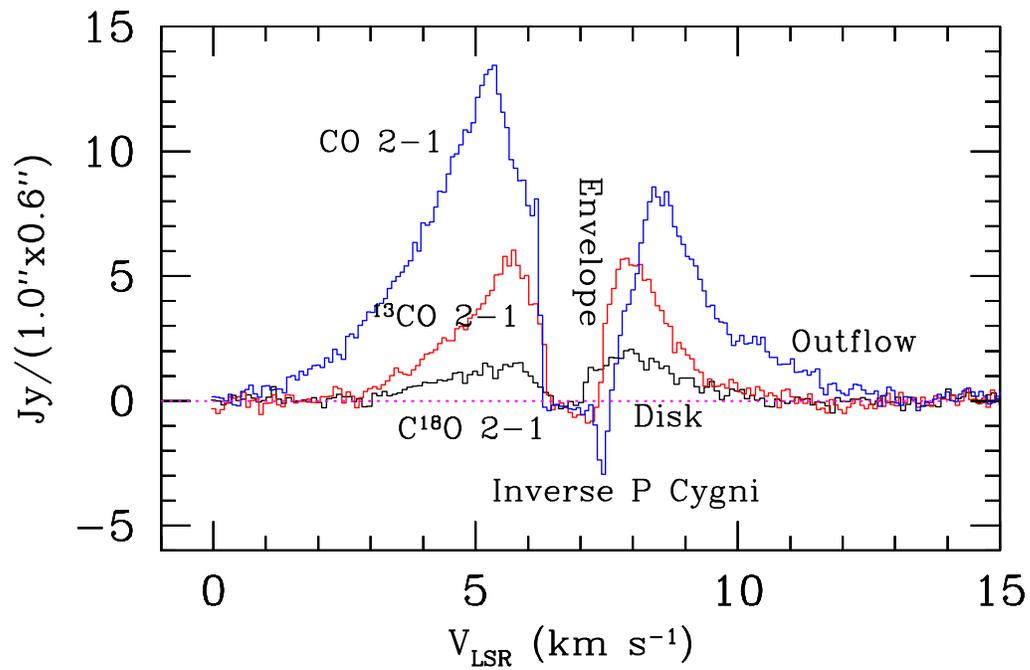}
\caption{
CO isotopologue line profiles integrated over the central area ($\sim 1\arcsec .0\times 0\arcsec.6$) of L328-IRS. These profiles show the signatures of  various kinematical motions  in this area,
outflow,  disk rotation, and infalling motions from envelope. }
\end{figure}

\clearpage
\begin{figure}
\centering
\includegraphics[height=6.0in,angle=0]{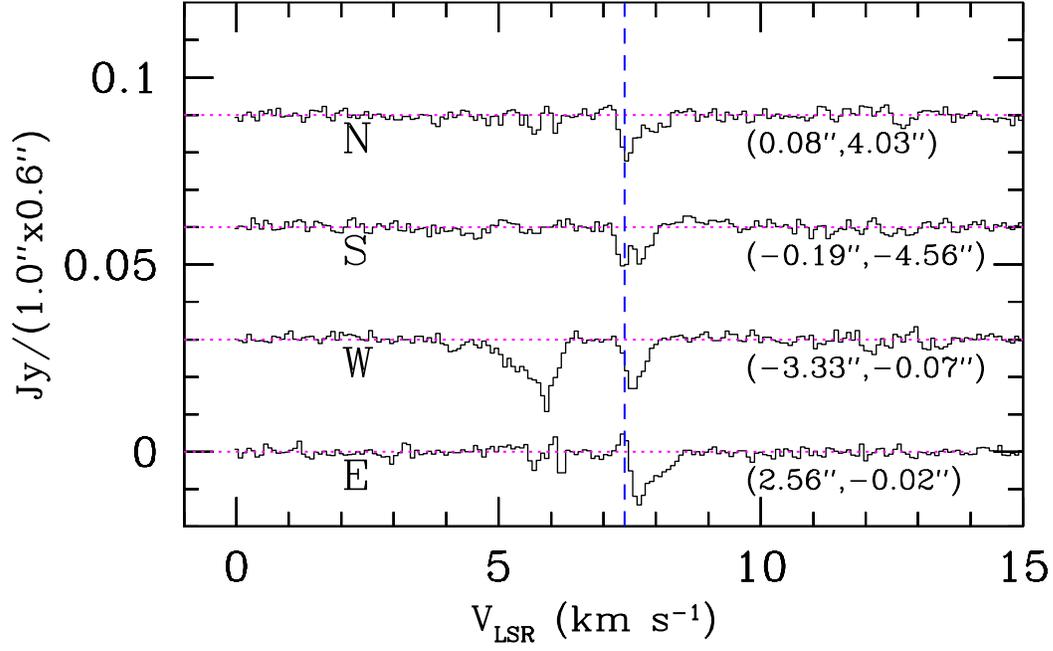}
\caption{
CO absorption profiles obtained (over the area of $~1\arcsec .0\times 0\arcsec.6$)  in the four directions  around L328-IRS. The directions N, S, W, and E,  and offset values are relatively given with respect to L328-IRS. A dashed line indicates the position of the LSR velocity 
($\rm \sim 7.4~km~s^{-1}$) for the absorption feature seen in the inverse P Cygni profile of CO in Figure 8. These absorption features seen at all the directions between $\rm \sim 7.4 - 8.0~km~s^{-1}$ are probably due to the absorption of 
the resolved-out background CO emission by the infalling cool gas in the foreground envelope. }
\end{figure}

\clearpage
\begin{figure}
\centering
\includegraphics[width=3.4in]{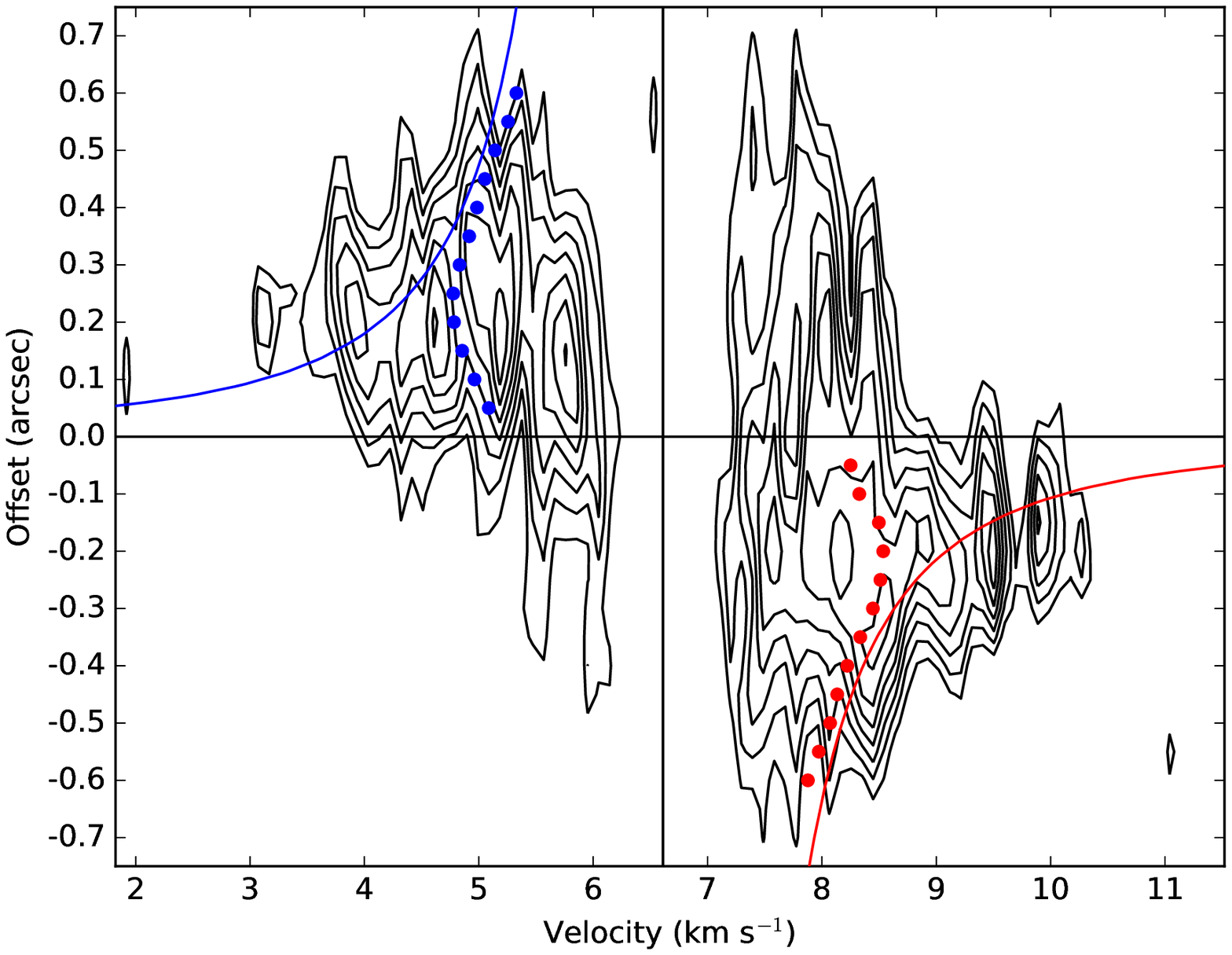}
\includegraphics[width=3.6in]{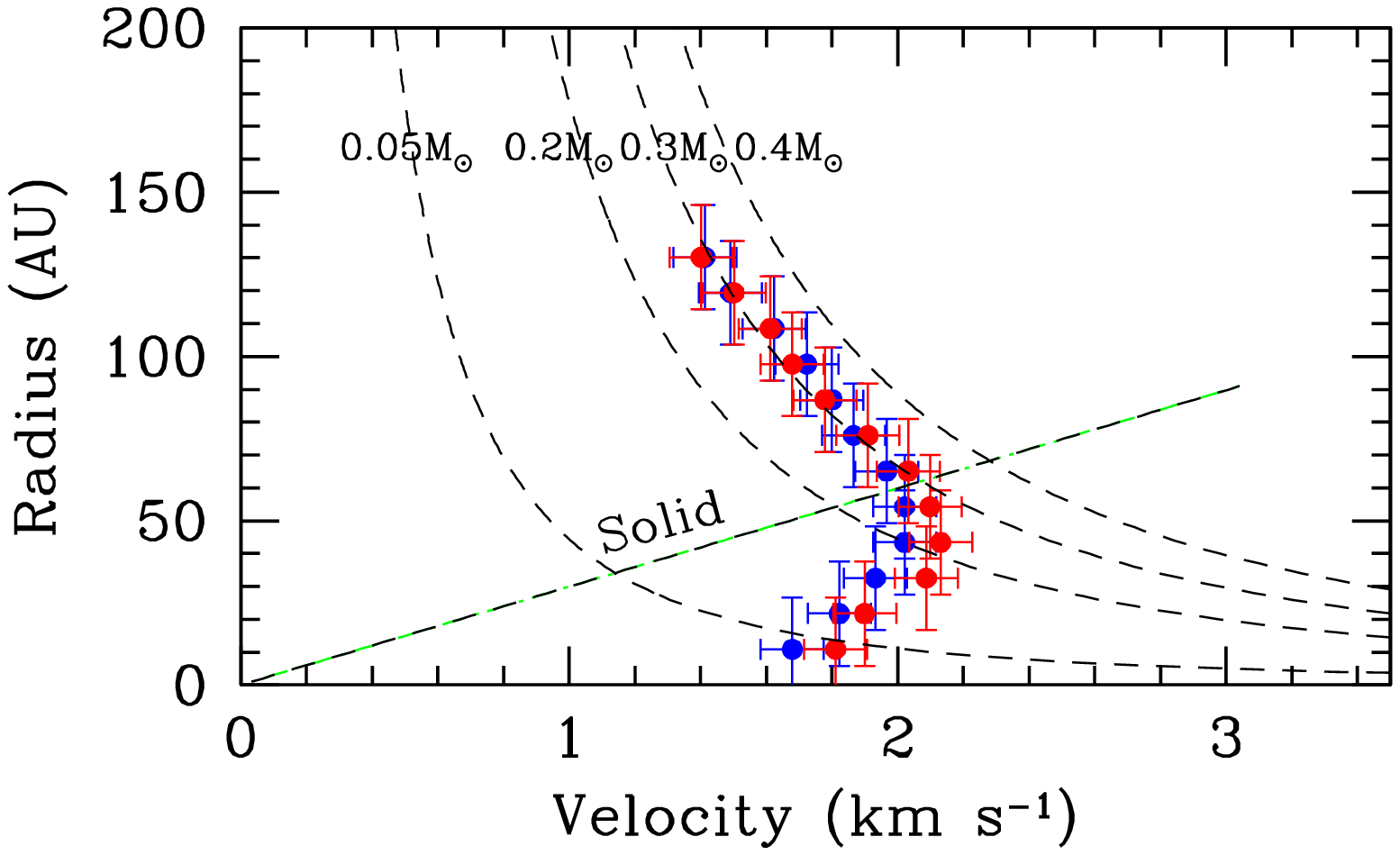}
\caption{Position-Velocity diagram of C$^{18}$O emission  in the disk (left panel) and  its rotation velocity (right panel).
Left panel shows a position velocity diagram of C$^{18}$O emission along the major axis of disk structure. Filled blue and red dots indicate the velocity points at the peak intensity obtained from Gaussian fit to the cut profile along the velocity axis at each offset position.
Blue and red curves are the trajectories that gases in the disk would rotate in  Keplerian motions by a central object of $\sim 0.3\msun$.
Right panel shows the rotational velocity as a function of the radius  obtained using PV diagrams of C$^{18}$O line data, indicating that the disk motions of L328-IRS are best fitted with Keplerian motions by a central object of $\sim 0.3\msun$ between 60 and 130 AU in radius 
while the motions within 60 AU are not fitted at all with Keplerian motions.}
\end{figure}

\clearpage
\begin{figure}
\centering
\includegraphics[width=3.4in]{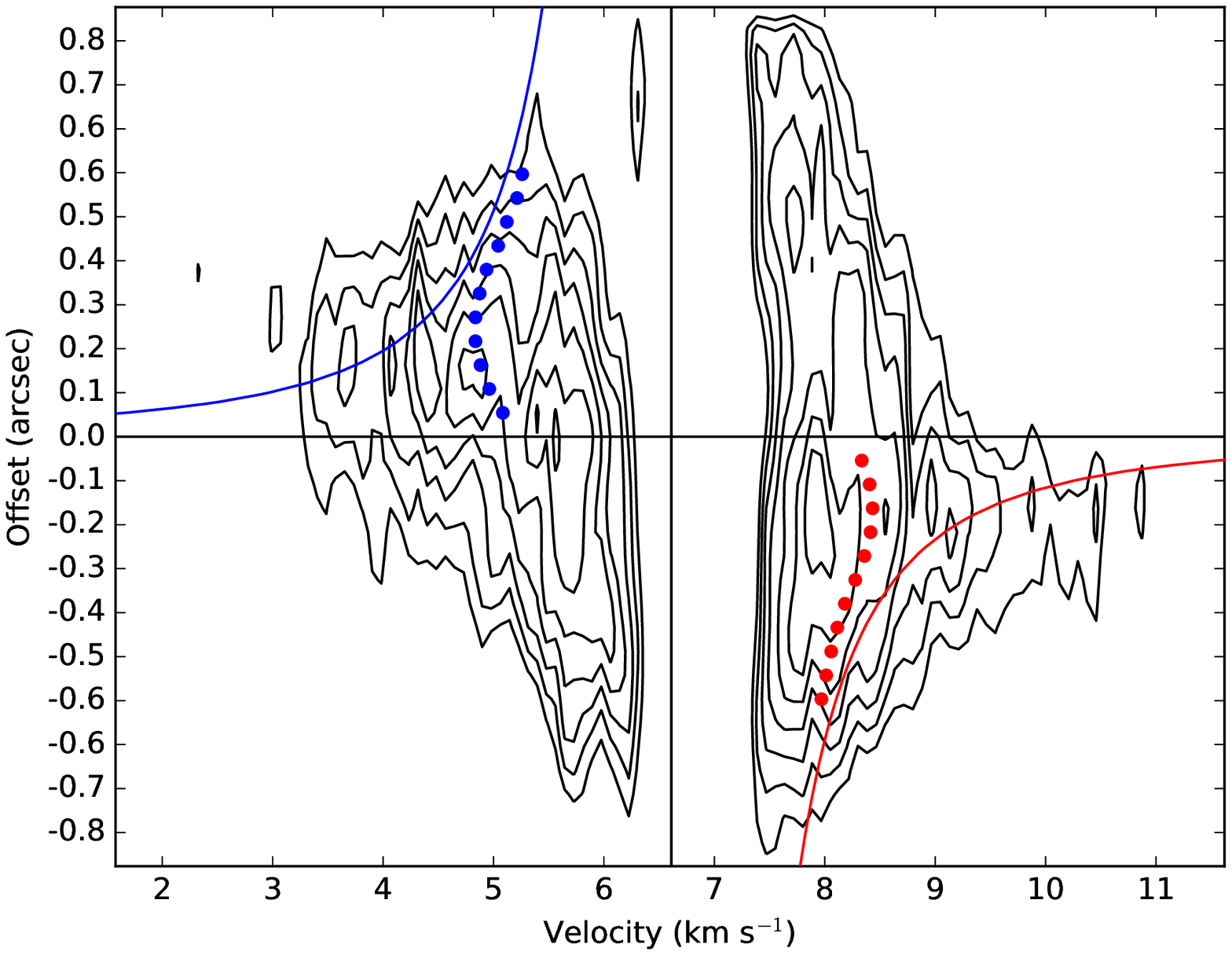}
\includegraphics[width=3.6in]{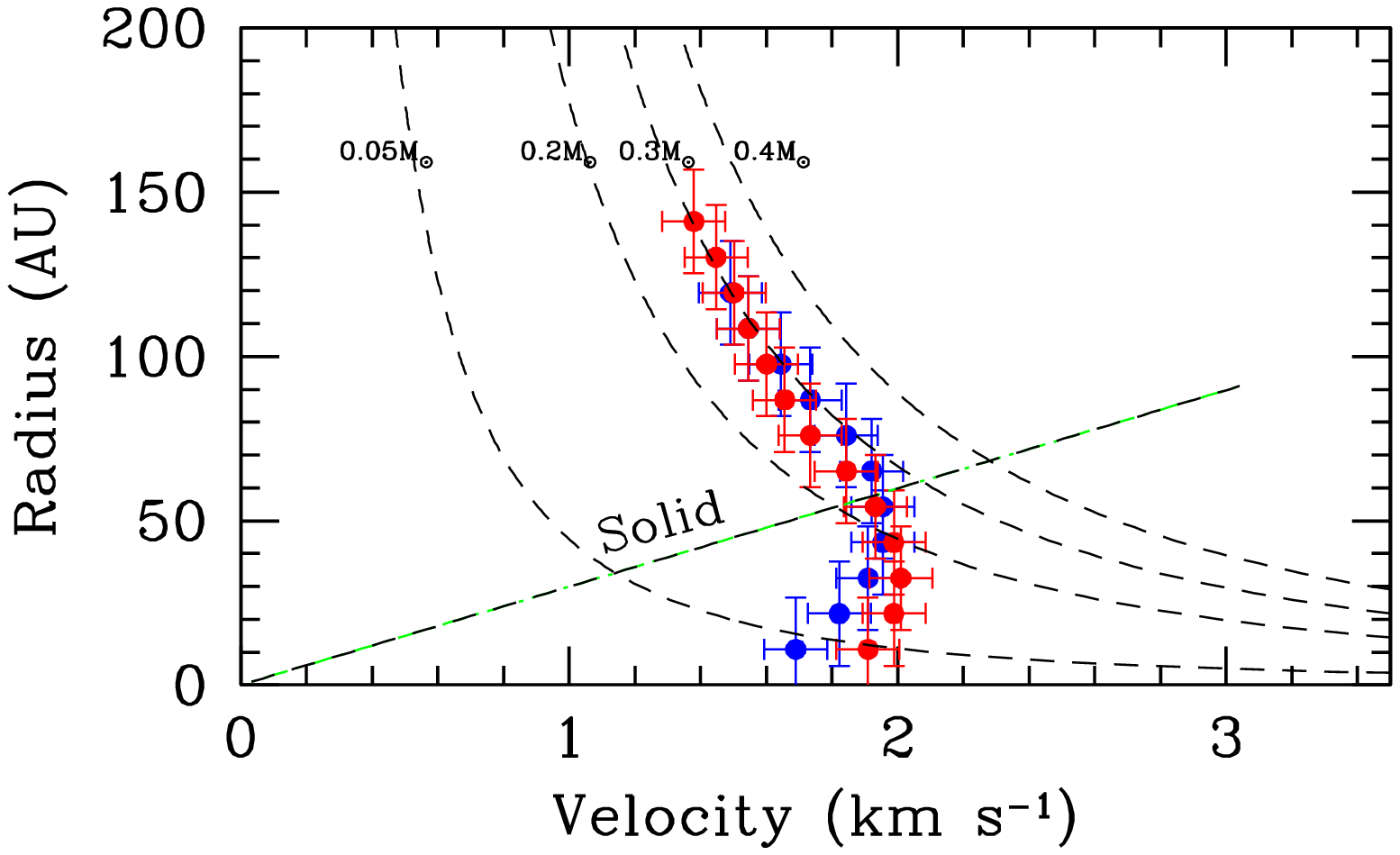}
\caption{
Position-Velocity diagram of $^{13}$CO emission  in the disk (left panel) and  its rotation velocity (right panel). All markers in the diagrams have the same meaning as those in Figure 10, except that $^{13}$CO line data are being used. The data  between 60 and 130 AU in radius in the diagrams are best fitted with Keplerian motions by a central object of $\sim 0.27\msun$ while the motions within 60 AU are not fitted at all with Keplerian motions.
}
\end{figure}

\clearpage
\begin{figure}
\centering
\includegraphics[width=6.0in]{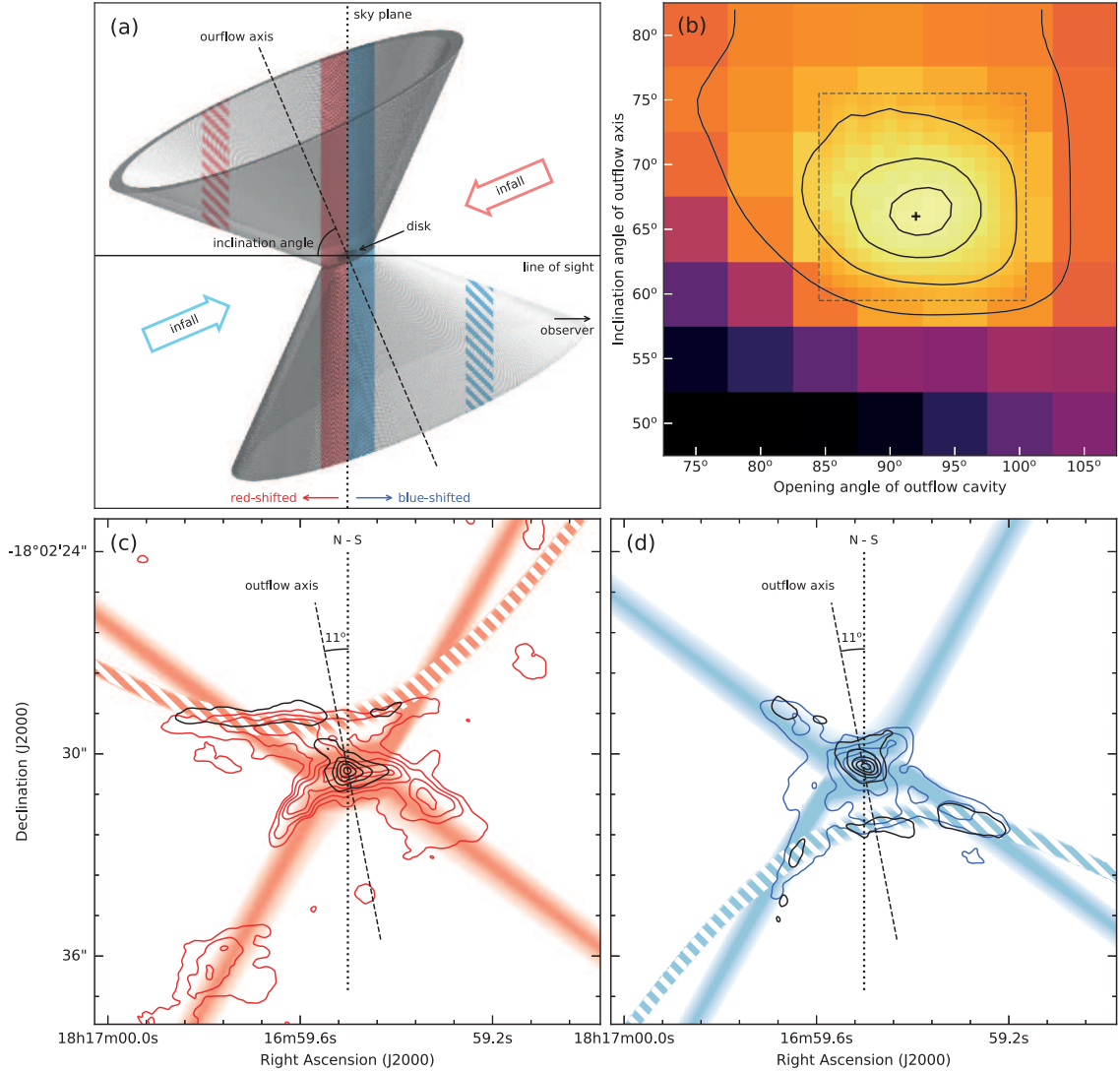}
\caption{Possible 3D shape of L328-IRS outflow system and comparison of its projected image with the observed image of L328-IRS region.
(a) Side view of our possible 3D model of L328-IRS system. Observer is looking at L328-IRS from right to left.  The dotted line from top to bottom indicates  the sky plane.
The conic structure has its opening angle of $92^{\circ}$ and inclination  angle of $66^{\circ}$, representing the shape of outflow cavities for L328-IRS. The simple velocity information for the conic outflows is implemented such that  the components nearer to observer with respect to the sky plane is blue-shifted and the components  farther from observer with respect to the sky plane is red-shifted. 
The regions painted in blue and red tones and slanted lines  in the outflow cavities are the parts used in reproducing observed features  in the intensity maps in (c) and (d).  
 (b) Intensity reproducibility of the observed images with our model L328-IRS system in space of two parameters, the opening angle of the outflow cavity and the inclination angle of outflow axis.
 The reproducibility is measured with a sum of the multiplied intensities between the observed data and  each model. The figure shows  the ratio for the sum value of each model with respect to the sum value of our  best  model with the opening angle of $92^{\circ}$ and inclination  angle of $66^{\circ}$. The ratio is expressed with the colour tones or contours in this panel. This was calculated at $5^{\circ}$ interval outside the dashed box and  at $1^{\circ}$ interval inside the dashed box. 
 The cross symbol indicates the point of the highest reproducibility, and contours are drawn in levels of 98, 94, 86, and 70 \% with respect to the cross point. 
 (c) Comparison of the red-shifted images (shown in contours) for the observed L328-IRS system  with those (drawn in red tone  and slanted lines) of our model L328-IRS system.
 Two overlaid contours in red and black contours are to show the integrated intensities between $\rm \sim 7.28 - 8.24~km~s^{-1}$   and $\rm \sim 8.32 - 10.40 ~km~s^{-1}$, respectively.  
 (d) Comparison of the blue-shifted images (shown in contours) for the observed L328-IRS system  with those (drawn in blue tone  and slanted lines) of our model L328-IRS system.
Two overlaid contours in blue and black contours are to show the integrated intensities between $\rm \sim 5.28 - 6.24~km~s^{-1}$   and $\rm \sim 2.24 - 5.20 ~km~s^{-1}$, respectively.  
 The projected images on the sky of this 3D configuration model of L328-IRS system  can mostly explain  well the observed features in details.  
}
\end{figure}

\end{document}